\documentclass{article} % For LaTeX2e
\usepackage{iclr2024_conference,times}

\usepackage{amsmath,amsfonts,bm}

\def\eqref#1{equation~\ref{#1}}
\def\1{\bm{1}}

\DeclareMathAlphabet{\mathsfit}{\encodingdefault}{\sfdefault}{m}{sl}
\SetMathAlphabet{\mathsfit}{bold}{\encodingdefault}{\sfdefault}{bx}{n}

\usepackage{booktabs}       % professional-quality tables
\usepackage{amsfonts}       % blackboard math symbols
\usepackage{nicefrac}       % compact symbols for 1/2, etc.
\usepackage{microtype}      % microtypography
\usepackage{xcolor}         % colors
\usepackage{amssymb}        % checkmark symbol
\usepackage{multirow}
\usepackage{graphicx}
\usepackage{colortbl}
\usepackage{pifont}
\usepackage{color,soul}
\usepackage{subfigure}
\usepackage{ragged2e}
\usepackage{hyperref}
\usepackage{url}
\newcommand{\crossmark}{\ding{55}}

\newcommand*\samethanks[1][\value{footnote}]{\footnotemark[#1]}

\defcitealias{diabetes2008use}{DirecNet, 2008}
\defcitealias{ada}{ADA, 2021}
\defcitealias{idf}{IDF, 2021}

\title{GlucoBench: Curated List of Continuous Glucose Monitoring Datasets with Prediction Benchmarks}
\iclrfinalcopy
\author{%
Renat Sergazinov$^{1}$\thanks{Address correspondence to: mrsergazinov@tamu.edu, irinag@tamu.edu}, Elizabeth Chun$^{1}$, Valeriya Rogovchenko$^{1}$, 
\\
\textbf{Nathaniel Fernandes$^{2}$,
Nicholas Kasman$^{2}$, Irina Gaynanova$^{1}$\samethanks[1]} \\
$^{1}$Department of Statistics, Texas A\&M University\\
$^{2}$Department of Electrical and Computer Engineering, Texas A\&M University%\\
}

\begin{document}

\maketitle

\begin{abstract}
    The rising rates of diabetes necessitate innovative methods for its management. Continuous glucose monitors (CGM) are small medical devices that measure blood glucose levels at regular intervals providing insights into daily patterns of glucose variation. 
    % Accurate prediction of glucose trajectories, based on CGM data, can significantly enhance diabetes management by  
    Forecasting of glucose trajectories based on CGM data  holds the potential to substantially improve diabetes management, by both refining artificial pancreas systems and enabling individuals to make adjustments based on  predictions to maintain optimal glycemic range.
    Despite numerous methods proposed for CGM-based glucose trajectory prediction, these methods are typically evaluated on small, private datasets, impeding reproducibility, further research, and practical adoption. 
    The absence of standardized prediction tasks and systematic comparisons between methods has led to uncoordinated research efforts, obstructing the identification of optimal tools for tackling specific challenges. 
    As a result, only a limited number of prediction methods have been implemented in clinical practice.  
    % Additionally, the lack of standardized prediction tasks and varying focus on CGM data aspects contribute to uncoordinated research efforts.

    To address these challenges, we present a comprehensive resource that provides (1) a consolidated repository of curated publicly available CGM datasets to foster reproducibility and accessibility; (2) a standardized task list to unify research objectives and facilitate coordinated efforts; (3) a set of benchmark models with established baseline performance, enabling the research community to objectively gauge new methods' efficacy; and (4) a detailed analysis of performance-influencing factors for model development. We anticipate these resources to propel collaborative research endeavors in the critical domain of CGM-based glucose predictions. {Our code is available online at github.com/IrinaStatsLab/GlucoBench.}
    % Our work is accessible at \verb|github.com/IrinaStatsLab/GlucoBench|.
        
\end{abstract}

\section{Introduction} %(Renat)

According to the International Diabetes Federation, 463 million adults worldwide have diabetes with 34.2 million people affected in the United States alone \citepalias{idf}.
% , and this number is expected to rise to 700 million by 2045 \citep{idf}. In the United States, 34.2 million people have diabetes, or 10.5\% of the population. 
Diabetes is a leading cause of heart disease \citep{nanayakkara2021impact}, blindness \citep{wykoff2021risk}, and kidney disease \citep{alicic2017diabetic}. Glucose management is a critical component of diabetes care, however achieving target glucose levels is difficult due to multiple factors that affect glucose fluctuations, e.g., diet, exercise, stress, medications, and individual physiological variations. % can significantly influence fluctuations in glycemic levels.

\begin{figure}[t]
    \centering
    \includegraphics[width=\textwidth]{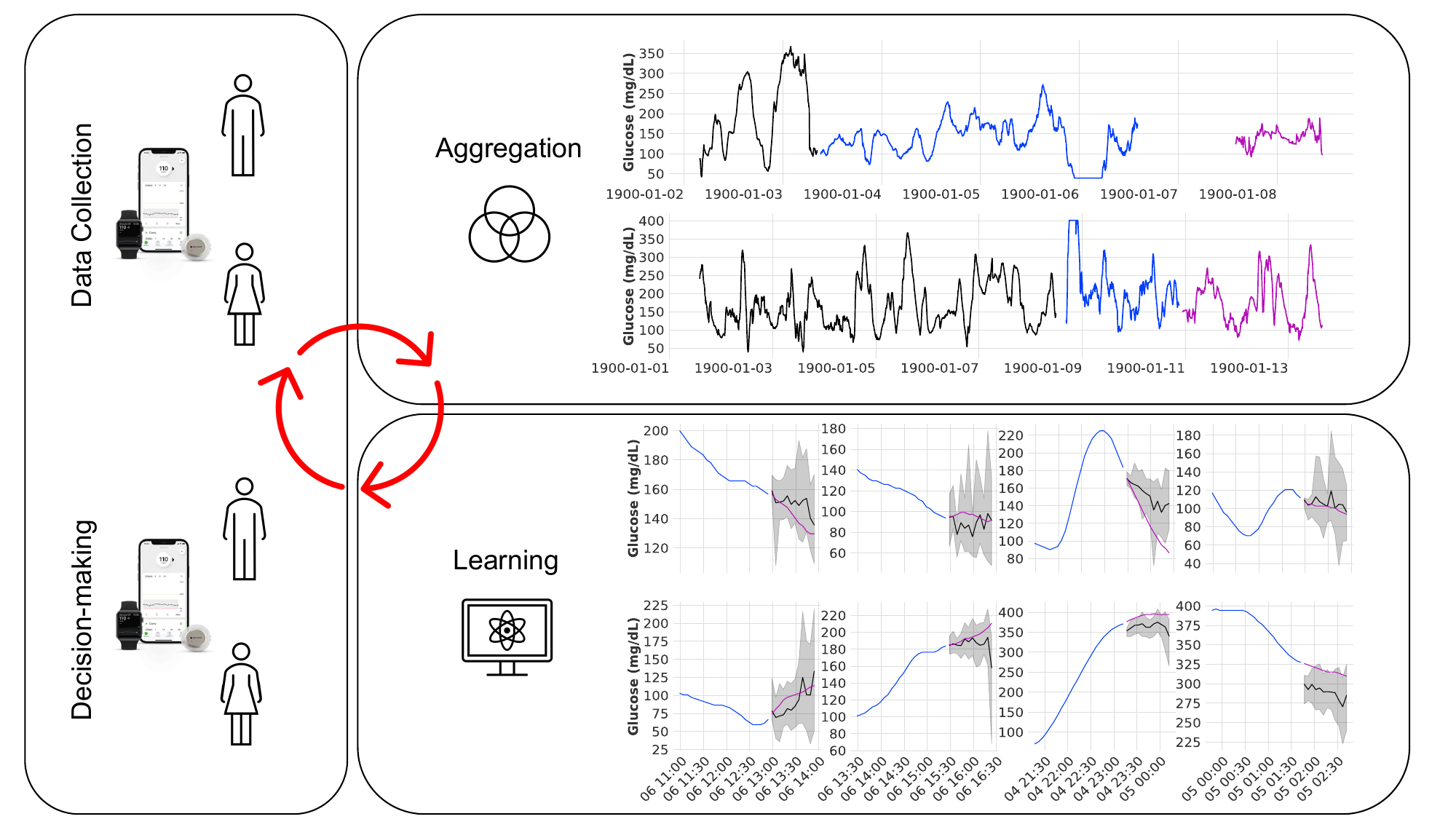}
    \caption{Sample of glucose curves captured by the Dexcom G4 Continuous Glucose Monitoring (CGM) system, with dates de-identified for privacy\citep{weinstock2016risk}.}
    \label{fig:glucose-prediction}
\end{figure}

Continuous glucose monitors (CGM) are medical devices that measure blood glucose levels at frequent intervals, often with a granularity of approximately one minute. CGMs have great potential to improve diabetes management by furnishing real-time feedback to patients and by enabling an autonomous artificial pancreas (AP) system when paired with an insulin pump \citep{contrerasArtificialIntelligenceDiabetes2018, kimLessonsUseContinuous2020}. Figure~\ref{fig:glucose-prediction} illustrates an example of a CGM-human feedback loop in a recommender setting. The full realization of CGM potential, however, requires accurate glucose prediction models.  Although numerous prediction models \citep{foxDeepMultiOutputForecasting2018a, armandpourDeepPersonalizedGlucose2021a, sergazinov2022gluformer} have been proposed, only simple physiological \citep{bergman1979quantitative, hovorka2004nonlinear} or statistical  \citep{oviedoReviewPersonalizedBlood2017, mirshekarianLSTMsNeuralAttention2019, xieBenchmarkingMachineLearning2020} models are utilized within current CGM and AP software. The absence of systematic model evaluation protocols and established benchmark datasets hinder the analysis of more complex models'  risks and benefits, leading to their limited practical adoption \citep{mirshekarianLSTMsNeuralAttention2019}.

In response, we present a curated list of five public CGM datasets and a systematic protocol for models' evaluation and benchmarking. The selected datasets have varying sizes and demographic characteristics, while the developed systematic data-preprocessing pipeline facilitates the inclusion of additional datasets. {We propose two tasks: (1) enhancing the predictive accuracy; (2) improving the uncertainty quantification (distributional fit) associated with predictions. In line with previous works \citep{mirshekarianLSTMsNeuralAttention2019, xieBenchmarkingMachineLearning2020}, we measure the performance on the first task with mean squared error (MSE) and mean absolute error (MAE), and on the second task with likelihood and expected calibration error (ECE) \citep{kuleshovCalibration2018}.} For each task, we train and evaluate a set of baseline models. From data-driven models, we select linear regression and ARIMA to represent shallow baselines, and Transformer \citep{vaswani2017attention}, NHiTS \citep{challuNhits2022n}, TFT \citep{lim2021temporal}, and Gluformer \citep{sergazinov2022gluformer} to represent deep learning baselines.  We select the Latent ODE \citep{rubanova2019latent} to represent a hybrid data-driven / physiological model. 
%Moreover, we provide an in-depth analysis of how factors such as distributional shifts in the data and the inclusion of covariates affect model performance

Our work contributes a \textbf{curated collection} of diverse CGM datasets, \textbf{formulation of two tasks} focused on model accuracy and uncertainty quantification, an efficient \textbf{benchmarking protocol}, evaluation of a range of \textbf{baseline models} including shallow, deep and hybrid methods, and a \textbf{detailed analysis of performance-influencing factors} for model optimization.

\section{Related works} %(Renat)
\label{sec:related-works}

\begin{table}[t]
    \caption{Summary of the glucose prediction models by dataset and model type. %For models, we denote "deep" for deep learning, "shallow" for other non-deep learning, and "physiological" for physiological models.
    We indicate "open" for datasets that are publicly available online, "simulation" for the ones based on simulated data, "proprietary" for the ones that cannot be released. We indicate deep learning models by "deep", non-deep learning models by "shallow", and physiological models by "physiological." We provide full table with references to all the works in Appendix A.}
    \label{table:related-works-short}
    \centering
            \small
    \begin{tabular}{cccccc}
        \toprule
        Type & Diabetes & \# of datasets & \# of deep & \# of shallow & \# of Physiological \\
        \midrule
        Open & Type 1 & 9 & 13 &  3 & 2\\
        Simulation & Type 1 & 12 & 3 & 3 & 6 \\
        Proprietary & Mixed & 22 & 7 & 8 & 7 \\
        \bottomrule
        
    \end{tabular}
\end{table}

An extensive review of glucose prediction models and their utility is provided by \citet{oviedoReviewPersonalizedBlood2017, contrerasArtificialIntelligenceDiabetes2018, kimLessonsUseContinuous2020, kavakiotisMachineLearningData2017}. Following \citet{oviedoReviewPersonalizedBlood2017}, we categorize prediction models as physiological, data-driven, or hybrid. \textbf{Physiological models}  rely on the mathematical formulation of the dynamics of insulin and glucose metabolism via differential equations \citep{man2014uva, lehmann1991physiological}. A significant limitation of these models is the necessity to pre-define numerous parameters. \textbf{Data-driven models} rely solely on the CGM data (and additional covariates when available) to characterize glucose trajectory without incorporating physiological dynamics. These models can be further subdivided into shallow (e.g. linear regression, ARIMA, random forest, etc.) and deep learning models (e.g. recurrent neural network models, Transformer, etc.). {Data-driven models, despite their capacity to capture complex patterns, may suffer from overfitting and lack interpretability.}
% They often require large labeled datasets and may overfit the training data, reducing their generalizability. Additionally, they often lack interpretability, making the understanding of their predictions challenging.}
\textbf{Hybrid models} use physiological models as a pre-processing or data augmentation tool for data-driven models. {Hybrid models enhance the flexibility of physiological models and facilitate the fitting process, albeit at the expense of diminished interpretability.} Table~\ref{table:related-works-short} summarizes existing models and datasets, indicating model type.

\textbf{Limitations.} The present state of the field is characterized by several key constraints, including (1) an absence of well-defined benchmark datasets and studies, 
% (2) a scarcity of benchmark studies comparing different models, 
(2) a dearth of open-source code bases, {and (3) omission of Type 2 diabetes from most open CGM studies}. To address the second limitation, two benchmark studies have been undertaken to assess the predictive performance of various models \citep{mirshekarianLSTMsNeuralAttention2019, xieBenchmarkingMachineLearning2020}. Nonetheless, these studies only evaluated the models on one dataset \citep{marling2020ohiot1dm}, comprising a limited sample of 5 patients with Type 1 diabetes, and failed to provide source code. We emphasize that, among the 45 methods identified in Table~\ref{table:related-works-short}, a staggering 38 works do not offer publicly available implementations. {For the limitation (3), it is important to recognize that Type 2 is more easily managed through lifestyle change and oral medications than Type 1 which requires lifelong insulin therapy.}
% pre-diabetes and 
% pre-diabetes can be delayed or even prevented from progressing to Type 2 with proper management, and 
% Therefore, better CGM forecasting methods are crucial for these groups of patients. For patients with pre-diabetes, forecasting insights may help to inform the lifestyle modification to prevent further progression of the disease. For patients with Type 2 diabetes, CGM forecasting could enhance glycemic control both through lifestyle change and timely oral medication and delay or reduce the reliance on the insulin intake, improving the overall life quality.

\section{Data}

\subsection{Description}

We have selected five publicly available CGM datasets: \citet{broll2021interpreting, colas2019detrended, dubosson2018open, hall2018glucotypes, weinstock2016risk}.

% We can highlight this ahole paragraph as red since it perfectly answers the question
{To ensure data quality, we used the following set of criteria.} First, we included a variety of dataset sizes and verified that each dataset has measurements on at least 5 subjects and that the collection includes a variety of sizes ranging from only five \citep{broll2021interpreting} to over 200 \citep{colas2019detrended, weinstock2016risk} patients.
{On the patient level}, we ensured that each subject has non-missing CGM measurements for at least 16 consecutive hours. {At the CGM curve level, we have verified that measurements fall within a clinically relevant range of 20 mg/dL to 400 mg/dL, avoiding drastic fluctuations exceeding 40 mg/dL within a 5-minute interval, and ensuring non-constant values.}

Finally, we ensured that the collection covers distinct population groups representing subjects with Type 1 diabetes \citep{dubosson2018open, weinstock2016risk}, Type 2 diabetes \citep{broll2021interpreting}, or a mix of Type 2 and none \citep{colas2019detrended, hall2018glucotypes}. We expect that the difficulty of accurate predictions will depend on the population group: patients with Type 1 have significantly larger and more frequent glucose fluctuatons.
% For example, we anticipate it would be easier to predict glucose values for healthy subjects who have less variable glucose patterns than subjects with Type I diabetes, who have significantly larger glucose fluctuations. 
%as Subject type is important because it heavily influences glucose variability and patterns. For example, healthy subjects are expected to have less variable glucose patterns, which we anticipate would be easier to predict. In contrast, subjects with type 1 diabetes tend to have large glucose fluctuations and are thus significantly more challenging to predict. %Finally, we include a variety of dataset sizes . %All datasets are publicly available although \citep{weinstock2016risk} has a data use agreement per the T1D exchange.
Table~\ref{demographics} summarizes all five datasets with associated demographic information, where some subjects are removed due to data quality issues as a result of pre-processing (Section \ref{pre-processing}). We describe data availability in Appendix~A.
% When Age and Sex information is not included with the data, it is extracted from the study description in the associated publication.  

\begin{table}[!t]
    \caption{Demographic information (average) for each dataset before (Raw) and after pre-processing (Processed). CGM indicates the device type; all devices have 5 minute measurment frequency.}
    \label{demographics}
    \centering
    \resizebox{\textwidth}{!}{%
    \begin{tabular}{ l | c c cc cc cc }
        \toprule
        Dataset & Diabetes & CGM & \multicolumn{2}{c}{\# of Subjects} & \multicolumn{2}{c}{Age} & \multicolumn{2}{c}{Sex (M / F)} \\ 
        \cmidrule(lr){2-2} \cmidrule{3-3} \cmidrule(lr){4-5} \cmidrule(lr){6-7} \cmidrule(lr){8-9}
        & Overall & Overall & Raw & Processed & Raw & Processed & Raw & Processed\\
        \midrule
        \citet{broll2021interpreting} &
            Type 2 & Dexcom G4 & 5 & 5 & NA & NA & NA & NA \\
        \citet{colas2019detrended}& 
             Mixed  & MiniMed iPro & 208 & 201 & 59 & 59 & 103 / 104 & 100 / 100 \\
        \citet{dubosson2018open} &
            Type 1 & MiniMed iPro2 & 9 & 7 & NA & NA & 6 / 3 & NA\\
        \citet{hall2018glucotypes}&
            Mixed & Dexcom G4 & 57 & 56 & 48 & 48 & 25 / 32 & NA\\
        \citet{weinstock2016risk}&
            Type 1 & Dexcom G4 & 200 & 192 & 68 & NA & 106 / 94 & 101 / 91 \\
        \bottomrule
    \end{tabular}
    }
\end{table}

 % \subsection{Features (Nicky)}
% This sections should include the description of the covariates we have for each dataset as well as our invented way to split the features into:
% \begin{enumerate}
%     \item Dynamic unknown covariates: covariates that change with time but are unknown in the future (for example heart rate);
%     \item Dynamic known covariates: covariates that change with time and that we know in the future (for example date, holidays);
%     \item Static covariates: covariates that do not change with time (marital status, annual income etc.).
% \end{enumerate}

 \textbf{Covariates.} In addition to CGM data, each dataset has covariates (features), which we categorize based on their temporal structure and input type.  The temporal structure distinguishes covariates as static (e.g. gender), dynamic known (e.g. hour, minute), and dynamic unknown (e.g. heart beat, blood pressure).  Furthermore, input types define covariates as either real-valued (e.g. age) or ordinal (e.g. education level) and categorical or unordered (e.g. gender) variables. We illustrate different types of temporal variables in Figure~\ref{fig:feature-categorization}. We summarize covariate types for each dataset in Appendix~A.

\begin{figure}[t]
    \centering
    \includegraphics[width=\linewidth]{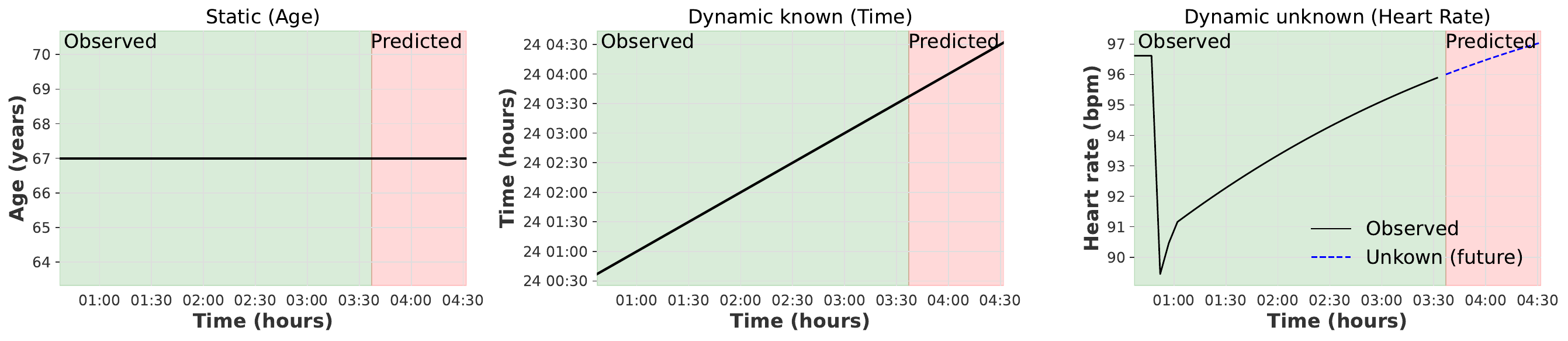}
    \caption{An illustration of static (Age), dynamic known (Date), and dynamic unknown (Heart Rate) covariate categories based on data from \citet{hall2018glucotypes} and \citet{dubosson2018open}.}\label{fig:feature-categorization}
\end{figure}

\subsection{Pre-processing} %(Valeriya)
\label{pre-processing}
% This section should include all steps we take from raw data to the model input:
% \begin{enumerate}
%     \item Describe the format that the CGM data comes in to us (be exact) \\

    % \item Encoding categorical features, mapping them to a numeric value (label encoder) \\

    % \item Interpolation (if less than a certain threshold, then we interpolate). Splitting into segments (if greater than some threshold, then we split into separate segments). \\

        % \item Splitting into train, validation, test, and out-of-distribution test \\

            % \item Min-max scaling to a range $[0, 1]$ (fit min-max scaler on the train part and transform all other sets) \\

We pre-process the raw CGM data via interpolation and segmentation, encoding categorical features, data partitioning, scaling, and division into input-output pairs for training a predictive model.

\begin{figure}[!t]
    \centering
    \includegraphics[width=0.73\textwidth]{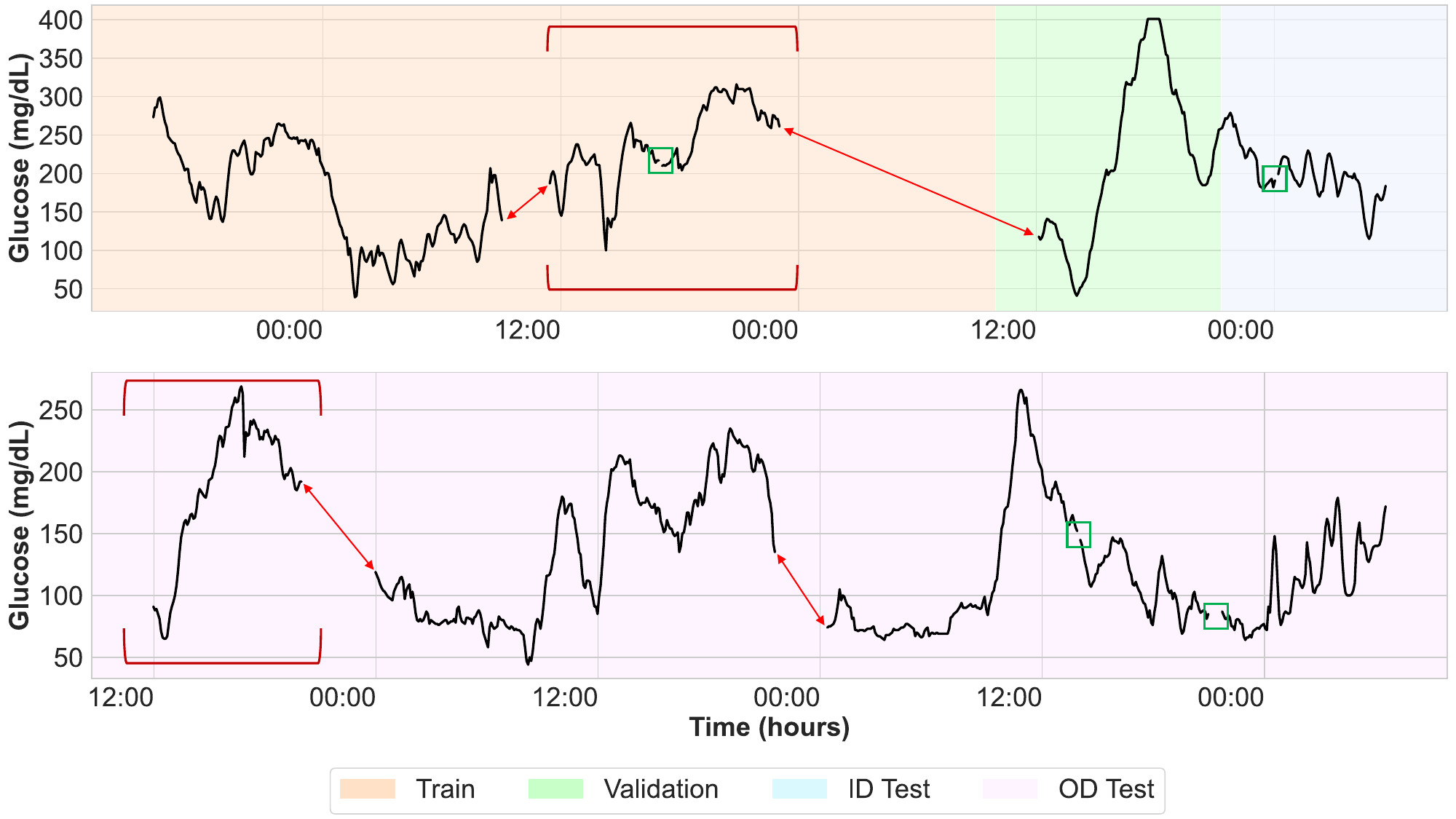}
    \caption{{Example data processing on \citet{weinstock2016risk}. The \textcolor{red}{red arrows} denote segmentation, \textcolor{green}{green blocks} interpolate values, and \textcolor{red}{red brackets} indicate dropped segments due to length.}}
    \label{fig:glucose-splitting}
\end{figure}

\textbf{Interpolation and segmentation.} To put glucose values on a uniform temporal grid, we identify gaps in  each subject's trajectory due to missing measurements. When the gap length is less than a predetermined threshold (Table~\ref{table:interpolation}), we impute the missing values by linear interpolation. When the gap length exceeds the threshold, we break the trajectory into several continuous segments.
Green squares in Figure~\ref{fig:glucose-splitting} indicate gaps that will be interpolated, whereas red arrows indicate larger gaps where the data will be broken into separate segments.
In \citet{dubosson2018open} dataset, we also interpolate missing values in dynamic covariates (e.g., heart rate). Thus, for each dataset we obtain a list of CGM sequences $\mathcal{D} = \{\mathbf{x}_{j}^{(i)}\}_{i , j}$ with $i$ indexing the patients and $j$ the continuous segments. Each segment $\mathbf{x}_{j}^{(i)}$ has length $L^{(i)}_j > L_{min}$, where $L_{min}$ is the pre-specified minimal value (Table~\ref{table:interpolation}).

\begin{table}[!t]
\caption{Interpolation parameters for datasets.}
\label{table:interpolation}
\centering
\begin{tabular}{ c | c c c c c }
\toprule
Parameters & Broll & Colas & Dubosson & Hall & Weinstock \\
\midrule
Gap threshold (minutes)  & 45 & 45 & 30 & 30 & 45              \\
Minimum length (hours) & 20 & 16 & 20 & 16 & 20   \\
\bottomrule
\end{tabular}
\end{table}

\textbf{Covariates Encoding.} While many of the covariates are real-valued, e.g., age, some covariates are categorical, e.g., sex. In particular, \citet{weinstock2016risk} dataset has 36 categorical covariates with an average of 10 levels per covariate. While one-hot encoding is a popular approach for modeling categorical covariates, it will lead to 360 feature columns on \citet{weinstock2016risk}, making it undesirable from model training time and memory perspectives. Instead, we use label encoding by converting each categorical level into a numerical value. Given $R$ covariates, we include them in the dataset as $\mathcal{D} = \{\mathbf{x}_{j}^{(i)}, \mathbf{c}_{1, j}^{(i)}, \dots \mathbf{c}_{R, j}^{(i)} \}_{i , j}$ where $\mathbf{c}_{r, j}^{(i)} \in \mathbb{R}$ for static and $\mathbf{c}_{r, j}^{(i)} \in \mathbb{R}^{L^{(i)}_j}$ for dynamic.

\textbf{Data splitting.} Each dataset is split into train, validation, and in-distribution (ID) test sets using 90\% of subjects. % as $\mathcal{D} = \mathcal{D}_{tr} \cup \mathcal{D}_{val} \cup \mathcal{D}_{id} \cup \mathcal{D}_{od}$ 
For each subject, the sets follow chronological time order as shown in Figure~\ref{fig:glucose-splitting}, with validation and ID test sets always being of a fixed length of 16 hours each (192 measurements). The data from the remaining 10\% of subjects is used to form an out-of-distribution (OD) test set to assess the generalization abilities of predictive models as in Section \ref{subsec:generalizability}. Thus, $\mathcal{D} = \mathcal{D}_{tr} \cup \mathcal{D}_{val} \cup \mathcal{D}_{id} \cup \mathcal{D}_{od}$.

%using the remaining 10\% of subjects. Most of the data is allocated to the training set for model development and optimization, while the validation and in-distribution (ID) test, each containing a fixed amount of 240 observations (20 hours) for each dataset, are used for hyperparameter fine-tuning, model selection, and performance evaluation. These datasets follow a chronological order, with the validation and in-distribution test sets succeeding the training set (as shown in the green and blue chart areas of Figure ng}). % By using both the ID and OD test sets, we can measure subject-specific trends and evaluate generalization, revealing potential challenges when deploying the model in clinical settings and informing efforts to enhance robustness.

\textbf{Scaling.} We use min-max scaling to standardize the measurement range for both glucose values and covariates. The minimum and maximum values are computed per dataset using $\mathcal{D}_{tr}$, and then the same values are used to rescale $\mathcal{D}_{val}$, $\mathcal{D}_{id}$, and $\mathcal{D}_{od}$. %Therefore, upon standardization, we have $\mathbf{x}_j^{(i)} \in [0, 1]^{L^{(i)}_j}$ and $\mathbf{c}_j^{(i)} \in [0, 1]^{L^{(i)}_j}$ or $\mathbf{c}_j^{(i)} \in [0, 1]$.

%proves to be a more appropriate approach considering the broad range of glucose levels across different subjects. Min-max scaling ensures that features are on a consistent scale, typically within the range of [0, 1]. We extract one minimum and maximum value across all segments and all subjects in the training set and then use it to scale all series in the train, validation, ID test, and OD test sets. By normalizing the data, the model can focus on learning the underlying patterns and relationships in glucose concentration, thereby improving its performance in predicting and analyzing glucose levels across various subjects and datasets.

%\textbf{Notation.} We represent each processed  We denote by 

\textbf{Input-output pairs.} Let $\mathbf{x}^{(i)}_{j, k:k+L}$ be a length $L$ contiguous slice of a segment from index $k$ to $k+L$. We define an input-output pair as $\{\mathbf{x}^{(i)}_{j, k : k+L}, \mathbf{y}^{(i)}_{j, k + L+1: k+L+T}\}$, where $\mathbf{y}^{(i)}_{j, k + L+1: k+L+T} = \mathbf{x}^{(i)}_{j, k + L+1: k+L+T}$ and $T$ is the length of prediction interval. Our choices of $T$, $L$ and $k$ are as follows. In line with the previous works \citep{oviedoReviewPersonalizedBlood2017}
    % \citep{foxDeepMultiOutputForecasting2018a, munoz-organeroDeepPhysiologicalModel2020, sergazinov2022gluformer, langaricaMetalearningApproachPersonalized2023},
    we focus on the 1-hour ahead forecasting ($T=12$ for 5 minute frequency). We treat $L$ as a hyper-parameter for model optimization since different models have different capabilities in capturing long-term dependencies. We sample $k$ without replacement from among the index set of the segment during training, similar to \citet{oreshkinNbeats2019n, challuNhits2022n}, and treat the total number of samples as a model hyper-parameter. We found the sampling approach to be preferable over the use of a sliding window with a unit stride \citep{dartsPackage2022}, as the latter is computationally prohibitive on larger training datasets and leads to high between-sample correlation, slowing convergence in optimization. We still use the sliding window when evaluating the model on the test set.

\section{Benchmarks}

\subsection{Tasks and metrics}

\textbf{Task 1: Predictive Accuracy.} Given the model prediction $\hat{\mathbf{y}}_{j, k + L: k+L+T }$, we measure accuracy on the test set using root mean squared error (RMSE) and mean absolute error (MAE):
$$
RMSE_{i,j,k} = \sqrt{\frac{1}{T} \sum_{t=1}^T \Big( y^{(i)}_{j, k + L + t} - \hat y^{(i)}_{j, k + L + t}\Big)^2}; \quad MAE_{i,j,k} = \frac{1}{T} \sum_{t=1}^T \Big |y^{(i)}_{j, k + L + t} - \hat y^{(i)}_{j, k + L + t}\Big |.
$$
Since the distribution of MAE and RMSE across samples is right-skewed, we use the median of the errors as has been done in \citet{sergazinov2022gluformer, armandpourDeepPersonalizedGlucose2021a}.

\textbf{Task 2: Uncertainty Quantification.} To measure the quality of uncertainty quantification, we use two metrics: log-likelihood on test data and calibration. For models that estimate a parametric predictive distribution over the future values, $\hat P^{(i)}_{j, k+L+1:k+L+T}:\mathbb{R}^{T}\to [0,1]$, we evaluate log-likelihood as
$$
\log L_{i, j, k} = \log \hat P^{(i)}_{j, k+L+1:k+L+T}\left (\mathbf{y}^{(i)}_{j, k+L+1:k+L+T} \right),
$$
where the parameters of the distribution are learned from training data, and the likelihood is evaluated on test data. Higher values indicate a better fit to the observed distribution. For both parametric and non-parametric models (such as quantile-based methods), we use regression calibration metric \citep{kuleshovCalibration2018}. The original metric is limited only to the univariate distributions.
% : (1) it only works with univariate distributions; (2) it assumes knowledge of $\hat F^{(i)}_{j,k+L+t}$, a marginal cumulative distribution function of the predictive distribution. 
To address the issue, we report an average calibration across marginal estimates for each time $t = 1, \dots, T$. 
% To address (2), we only utilize calibration with parametric models (for which exact analytical expression is closed-form) and quantile-based models (which directly estimate distribution function). 
To compute marginal calibration at time $t$, we (1) pick $M$ target confidence levels $0<p_1 < \dots < p_M<1$; (2)~estimate realized confidence level $\hat p_m$ using $N$ test input-output pairs as
%\begin{enumerate}
 %   \item Choose $M$ confidence levels, $p_1 < \dots < p_M$
%    \item Estimate 
    $$\hat p_m = \frac{\left|\Big\{y^{(i)}_{j, k + L + t} | \hat F^{(i)}_{j,k+L+t}\Big(y^{(i)}_{j, k + L + t}\Big) \leq p_m\Big\}\right|}{N};$$
    and (3) compute calibration across all $M$ levels as
    $$
    Cal_t = \sum_1^M (p_m - \hat p_m)^2.
    $$
%\end{enumerate}
The smaller the calibration value, the better the match between the estimated and true levels.

\subsection{Models}
To benchmark the performance on the two tasks, we compare the following models. \textbf{ARIMA} is a classical time-series model, which has been previously used for glucose predictions \citep{otoomRealTimeStatisticalModeling2015, yangARIMAModelAdaptive2019}. \textbf{Linear regression} is a simple baseline with a separate model for each time step $t = 1, \dots, T$. \textbf{XGBoost} \citep{chen2016xgboost} is gradient-boosted tree method, with a separate model for each time step $t$ to support multi-output regression.
\textbf{Transformer} represents a standard encoder-decoder auto-regressive Transformer implementation \citep{vaswani2017attention}. \textbf{Temporal Fusion Transformer (TFT)} is a quantile-based model that uses RNN with attention. TFT is the only model that offers out-of-the-box support for static, dynamic known, and dynamic unknown covariates. \textbf{NHiTS} uses neural hierarchical interpolation for time series, focusing on the frequency domain \citep{challuNhits2022n}. \textbf{Latent ODE} uses a recurrent neural network (RNN) to encode the sequence to a latent representation \citep{rubanova2019latent}. The dynamics in the latent space are captured with another RNN with hidden state transitions modeled as an ODE. Finally, a generative model maps the latent space back to the original space. \textbf{Gluformer} is a probabilistic Transformer model that models forecasts using a mixture distribution \citep{sergazinov2022gluformer}. 
For ARIMA, we use the code from \citep{garza2022statsforecast} which implements the algorithm from \citep{hyndmanAutoARIMA2008}.  For linear regression, XGBoost, TFT, and NHiTS, we use the open-source DARTS library \citep{dartsPackage2022}. For Latent ODE and Gluformer, we use the implementation in PyTorch \citep{rubanova2019latent, sergazinov2022gluformer}. We report the compute resources in Appendix~C. 

\subsection{Testing protocols} %(Nicky)
\label{subsec:testing-protocol}

In devising the experiments, we pursue the principles of reproducibility and fairness to all methods. 

\textbf{Reproducibility.} As the performance results are data split dependent, we train and evaluate each model using the same two random splits.  Additionally, all stochastically-trained models (tree-based and deep learning) are initialized 10 times on each training set with different random seeds.  Thus, each stochastically-trained model is re-trained and re-evaluated 20 times, and each deterministically-trained model 2 times, with the final performance score taken as an average across evaluations. We report standard error of each metric across the re-runs in Appendix B.

\textbf{Fairness.} To promote fairness and limit the scope of comparisons, we focus on out-of-the-box model performance when establishing baselines.  Thus, we do not consider additional model-specific tuning that could lead to performance improvements, e.g.,  pre-training, additional loss functions, data augmentation, distillation, learning rate warm-up, learning rate decay, etc. However, since model hyper-parameters can significantly affect performance, we automate the selection of these parameters. For ARIMA, we use the native automatic hyper-parameter selection algorithm provided in \citep{hyndmanAutoARIMA2008}. For all other models, we use Optuna \citep{akiba2019optuna} to run Bayesian optimization 
% (Tree-structured Parzen Estimator) 
with a fixed budget of 50 iterations. {We provide a discussion on the selected optimal model hyperparameters for each dataset in the supplement (Appendix~C).}

\subsection{Results} %(Lizzie)
We trained and tested each model outlined above on all five datasets using the established protocols. Table~\ref{table:results-short-acc} 
% and \ref{table:results-short-uq} 
present the results for the best-performing models on Task 1 (predictive accuracy) and Task~2 (uncertainty quantification). Appendix~B includes full tables for all models {together with standard errors and the visualized forecasts for the best models on \citep{weinstock2016risk} dataset.}
% We report our results on the first task in Table~\ref{table:results-task1} and on second task in Table~\ref{table:results-task2}.

On Task 1, the simple ARIMA and linear regression models have the highest accuracy on all but two datasets. On \citet{hall2018glucotypes} dataset ({mixed subjects including normoglycemic, prediabetes and Type 2 diabetes}), the Latent ODE model performs the best. On \citet{weinstock2016risk} dataset (the largest dataset), the Transformer model performs the best.

On Task 2, Gluformer model achieves superior performance as measured by model likelihood on all datasets. In regards to calibration, Gluformer is best on all but two datasets. On \citet{colas2019detrended} and \citet{weinstock2016risk} datasets (the largest datasets), the best calibration is achieved by TFT.

\section{Analysis}
\begin{figure}[t]
    \centering
    \includegraphics[width=\linewidth]{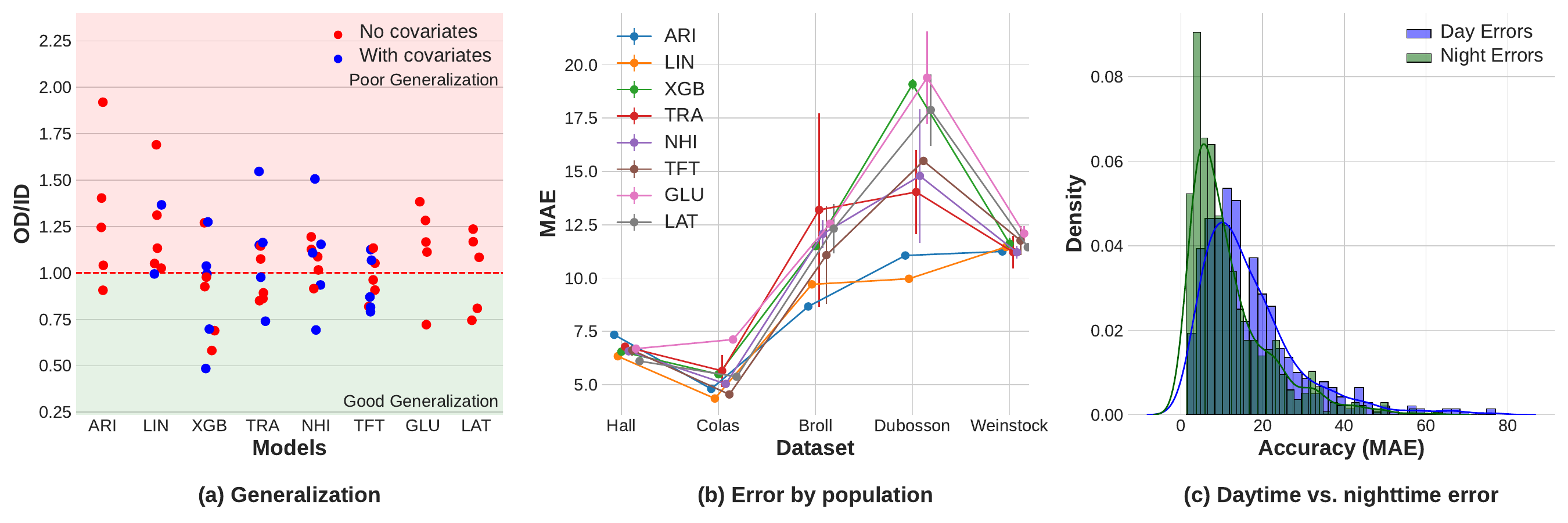}
    \caption{Analysis of errors by: (a) OD versus ID, (b) population diabetic type (healthy $\rightarrow$ Type 2 $\rightarrow$ Type 1), (c) daytime (9:00AM to 9:00PM) versus nighttime (9:00PM to 9:00AM).}
    \label{fig:analysis}
\end{figure}

\label{sec:analysis}
\begin{table}[t]
\centering
\caption{Accuracy and uncertainty metrics for selected models based on in-distribution (ID) test set without covariates. The selected models are best on at least one dataset for at least one metric. The best results on each data set are highlighted in \textbf{boldface}. TFT lacks likelihood information as it is a quantile-based model. Standard errors are reported in Appendix B.}
\label{table:results-short-acc}
\resizebox{\textwidth}{!}{%
\begin{tabular}{c|cccccccccccc}
\toprule
\multirow{2}{*}{Accuracy} & 
 \multicolumn{2}{c}{Broll} & \multicolumn{2}{c}{Colas} & \multicolumn{2}{c}{Dubosson} & \multicolumn{2}{c}{Hall} & \multicolumn{2}{c}{Weinstock}  \\ 
\cmidrule(lr){2-3} \cmidrule(lr){4-5} \cmidrule(lr){6-7} \cmidrule(lr){8-9} \cmidrule(lr){10-11}
& \cellcolor{lightgray} RMSE & \cellcolor{lightgray}MAE & \cellcolor{lightgray} RMSE & \cellcolor{lightgray}MAE & \cellcolor{lightgray} RMSE & \cellcolor{lightgray}MAE & \cellcolor{lightgray} RMSE & \cellcolor{lightgray}MAE & \cellcolor{lightgray} RMSE & \cellcolor{lightgray}MAE \\
\midrule
ARIMA & \textbf{10.53}&\textbf{8.67}&5.80&4.80&13.53&11.06&8.63&7.34&13.40&11.25 \\
Linear  & 11.68&9.71&\textbf{5.26}&\textbf{4.35}&\textbf{12.07}&\textbf{9.97}&7.38&6.33&13.60&11.46 \\
Latent ODE  & 14.37&12.32&6.28&5.37&20.14&17.88&\textbf{7.13}&\textbf{6.11}&13.54&11.45 \\
Transformer  & 15.12&13.20&6.47&5.65&16.62&14.04&7.89&6.78&\textbf{13.22}&\textbf{11.22} \\
\midrule
Uncertainty & \cellcolor{lightgray} Lik. & \cellcolor{lightgray} Cal. & \cellcolor{lightgray} Lik. & \cellcolor{lightgray} Cal. & \cellcolor{lightgray} Lik. & \cellcolor{lightgray} Cal. & \cellcolor{lightgray} Lik. & \cellcolor{lightgray} Cal. & \cellcolor{lightgray} Lik. & \cellcolor{lightgray} Cal. \\
\midrule
Gluformer & \textbf{-2.11}&\textbf{0.05}&\textbf{-1.07}&0.14&\textbf{-2.15}&\textbf{0.06}&\textbf{-1.56}&\textbf{0.05}&\textbf{-2.50}&0.08
\\
TFT & --&0.16&--&\textbf{0.07}&--&0.23&--&0.07&--&\textbf{0.07}
\\
\bottomrule
\end{tabular}
}
\end{table}

% \begin{table}[t]
% \centering
% \caption{Uncertainty quantification metrics for selected models based on in-distribution (ID) test set without covariates. The selected models are best on at least one dataset for at least one metric. The best results on each data set are highlighted in \textbf{boldface}. TFT lacks likelihood information as it is a quantile-based model. Standard errors are reported in Appendix B.}
% \label{table:results-short-uq}
% \resizebox{\textwidth}{!}{%
% \begin{tabular}{c|cccccccccccc}
% \toprule
% Uncertainty & \cellcolor{lightgray} Lik. & \cellcolor{lightgray} Cal. & \cellcolor{lightgray} Lik. & \cellcolor{lightgray} Cal. & \cellcolor{lightgray} Lik. & \cellcolor{lightgray} Cal. & \cellcolor{lightgray} Lik. & \cellcolor{lightgray} Cal. & \cellcolor{lightgray} Lik. & \cellcolor{lightgray} Cal. \\
% \midrule
% Gluformer & \textbf{-2.11}&\textbf{0.05}&\textbf{-1.07}&0.14&\textbf{-2.15}&\textbf{0.06}&\textbf{-1.56}&\textbf{0.05}&\textbf{-2.50}&0.08
% \\
% TFT & --&0.16&--&\textbf{0.07}&--&0.23&--&0.07&--&\textbf{0.07}
% \\
% \bottomrule
% \end{tabular}
% }
% \end{table}
\begin{table}[t]
\centering
\caption{Change in accuracy and uncertainty tasks between ID and OD sets. We indicate increases in performance in \textcolor{blue}{blue} and decreases in \textcolor{red}{red}. TFT lacks likelihood information as it is a quantile-based model.}
\label{table:results-short-acc-od}
\resizebox{\textwidth}{!}{%
\begin{tabular}{c|cccccccccccc}
\toprule
\multirow{2}{*}{Accuracy} & 
 \multicolumn{2}{c}{Broll} & \multicolumn{2}{c}{Colas} & \multicolumn{2}{c}{Dubosson} & \multicolumn{2}{c}{Hall} & \multicolumn{2}{c}{Weinstock}  \\ 
\cmidrule(lr){2-3} \cmidrule(lr){4-5} \cmidrule(lr){6-7} \cmidrule(lr){8-9} \cmidrule(lr){10-11}
& \cellcolor{lightgray} RMSE & \cellcolor{lightgray}MAE & \cellcolor{lightgray} RMSE & \cellcolor{lightgray}MAE & \cellcolor{lightgray} RMSE & \cellcolor{lightgray}MAE & \cellcolor{lightgray} RMSE & \cellcolor{lightgray}MAE & \cellcolor{lightgray} RMSE & \cellcolor{lightgray}MAE \\
\midrule
ARIMA & \textcolor{red}{+11.59\%}& \textcolor{red}{+12.01\%}& \textcolor{red}{+2.02\%}& \textcolor{red}{+1.49\%}& \textcolor{red}{+38.56\%}& \textcolor{red}{+31.81\%}&  \textcolor{blue}{-4.79\%}&  \textcolor{blue}{-5.08\%}& \textcolor{red}{+18.47\%}& \textcolor{red}{+18.56\%} \\
Linear  & \textcolor{red}{+2.51\%} & \textcolor{red}{+1.29\%} & \textcolor{red}{+1.27\%} & \textcolor{red}{+1.38\%} & \textcolor{red}{+30.01\%} & \textcolor{red}{+19.41\%} & \textcolor{red}{+6.46\%} & \textcolor{red}{+4.58\%} & \textcolor{red}{+14.5\%} & \textcolor{red}{+14.22\%} \\
Latent ODE  & \textcolor{red}{+4.12\%} & \textcolor{red}{+5.93\%} &  \textcolor{blue}{-10.05\%} &  \textcolor{blue}{-9.83\%} &  \textcolor{blue}{-13.7\%} &  \textcolor{blue}{-15.47\%} & \textcolor{red}{+8.07\%} & \textcolor{red}{+8.18\%} & \textcolor{red}{+11.17\%} & \textcolor{red}{+11.08\%}  \\
Transformer  & \textcolor{blue}{-7.16\%} &  \textcolor{blue}{-6.96\%} &  \textcolor{blue}{-7.78\%} &  \textcolor{blue}{-7.23\%} &  \textcolor{blue}{-5.52\%} &  \textcolor{blue}{-7.52\%} & \textcolor{red}{+3.69\%} & \textcolor{red}{+4.15\%} & \textcolor{red}{+7.0\%} & \textcolor{red}{+6.16\%}
\\
\midrule
Uncertainty & \cellcolor{lightgray} Lik. & \cellcolor{lightgray} Cal. & \cellcolor{lightgray} Lik. & \cellcolor{lightgray} Cal. & \cellcolor{lightgray} Lik. & \cellcolor{lightgray} Cal. & \cellcolor{lightgray} Lik. & \cellcolor{lightgray} Cal. & \cellcolor{lightgray} Lik. & \cellcolor{lightgray} Cal. \\
\midrule
Gluformer &
\textcolor{blue}{+6.72\%} & \textcolor{red}{+106.76\%} & \textcolor{red}{-50.33\%} & \textcolor{blue}{-29.83\%} & \textcolor{blue}{+45.64\%} & \textcolor{red}{+83.63\%} & \textcolor{blue}{+7.69\%} & \textcolor{red}{+9.24\%} & \textcolor{blue}{+3.33\%} & \textcolor{red}{+6.23\%}
\\
TFT & -- & \textcolor{blue}{-4.8\%} & -- & \textcolor{red}{+15.18\%} & -- & \textcolor{red}{+10.56\%} & -- & \textcolor{red}{+30.52\%} & -- & \textcolor{red}{+9.94\%}
\\
\bottomrule
\end{tabular}
}
\end{table}

% \begin{table}[t]
% \centering
% \caption{Change in uncertainty quantification between ID and OD sets. We indicate increases in performance in \textcolor{blue}{blue} and decreases in \textcolor{red}{red}. TFT lacks likelihood information as it is a quantile-based model.}
% \label{table:results-short-uq-od}
% \resizebox{\textwidth}{!}{%
% \begin{tabular}{c|cccccccccccc}
% \toprule

% Uncertainty & \cellcolor{lightgray} Lik. & \cellcolor{lightgray} Cal. & \cellcolor{lightgray} Lik. & \cellcolor{lightgray} Cal. & \cellcolor{lightgray} Lik. & \cellcolor{lightgray} Cal. & \cellcolor{lightgray} Lik. & \cellcolor{lightgray} Cal. & \cellcolor{lightgray} Lik. & \cellcolor{lightgray} Cal. \\
% \midrule
% Gluformer &
% \textcolor{blue}{+6.72\%} & \textcolor{red}{+106.76\%} & \textcolor{red}{-50.33\%} & \textcolor{blue}{-29.83\%} & \textcolor{blue}{+45.64\%} & \textcolor{red}{+83.63\%} & \textcolor{blue}{+7.69\%} & \textcolor{red}{+9.24\%} & \textcolor{blue}{+3.33\%} & \textcolor{red}{+6.23\%}
% \\
% TFT & -- & \textcolor{blue}{-4.8\%} & -- & \textcolor{red}{+15.18\%} & -- & \textcolor{red}{+10.56\%} & -- & \textcolor{red}{+30.52\%} & -- & \textcolor{red}{+9.94\%}
% \\
% \bottomrule
% \end{tabular}
% }
% \end{table}
\begin{table}[t]
\centering
\caption{Changes in accuracy and uncertainty tasks with and without covariates on ID test set. We indicate increases in performance in \textcolor{blue}{blue} and decreases in \textcolor{red}{red}. TFT lacks likelihood information as it is a quantile-based model.}
\label{table:results-short-acc-cov}
\resizebox{\textwidth}{!}{%
\begin{tabular}{c|cccccccccccc}
\toprule
\multirow{2}{*}{Accuracy} & 
 \multicolumn{2}{c}{Broll} & \multicolumn{2}{c}{Colas} & \multicolumn{2}{c}{Dubosson} & \multicolumn{2}{c}{Hall} & \multicolumn{2}{c}{Weinstock}  \\ 
\cmidrule(lr){2-3} \cmidrule(lr){4-5} \cmidrule(lr){6-7} \cmidrule(lr){8-9} \cmidrule(lr){10-11}
& \cellcolor{lightgray} RMSE & \cellcolor{lightgray}MAE & \cellcolor{lightgray} RMSE & \cellcolor{lightgray}MAE & \cellcolor{lightgray} RMSE & \cellcolor{lightgray}MAE & \cellcolor{lightgray} RMSE & \cellcolor{lightgray}MAE & \cellcolor{lightgray} RMSE & \cellcolor{lightgray}MAE \\
\midrule
Linear  & \textcolor{blue}{-14.82\%} & \textcolor{blue}{-13.34\%} & \textcolor{red}{+5.54\%} & \textcolor{red}{+5.75\%} & \textcolor{red}{+2.84\%} & \textcolor{red}{+0.61\%} & \textcolor{red}{+6.17\%} & \textcolor{red}{+5.09\%} & \textcolor{blue}{-1.54\%} & \textcolor{blue}{-1.08\%}  \\
Transformer  & \textcolor{blue}{-15.14\%} & \textcolor{blue}{-14.64\%} & \textcolor{red}{+30.31\%} & \textcolor{red}{+37.56\%} & \textcolor{red}{+64.99\%} & \textcolor{red}{+73.82\%} & \textcolor{blue}{-5.06\%} & \textcolor{blue}{-5.4\%} & \textcolor{red}{+9.33\%} & \textcolor{red}{+12.41\%}
\\
\midrule
Uncertainty & \cellcolor{lightgray} Lik. & \cellcolor{lightgray} Cal. & \cellcolor{lightgray} Lik. & \cellcolor{lightgray} Cal. & \cellcolor{lightgray} Lik. & \cellcolor{lightgray} Cal. & \cellcolor{lightgray} Lik. & \cellcolor{lightgray} Cal. & \cellcolor{lightgray} Lik. & \cellcolor{lightgray} Cal. \\
\midrule 
TFT & -- & \textcolor{red}{+94.6\%} & -- & \textcolor{red}{+114.61\%} & -- & \textcolor{red}{+7.57\%} & -- & \textcolor{red}{+16.84\%} & -- & \textcolor{blue}{-21.55\%}
\\
\bottomrule
\end{tabular}
}
\end{table}

% \begin{table}[t]
% \centering
% \caption{Changes in uncertainty quantification with and without covariates on ID test set. We indicate increases in performance in \textcolor{blue}{blue} and decreases in \textcolor{red}{red}. TFT lacks likelihood information as it is a quantile-based model.}
% \label{table:results-short-uq-cov}
% \resizebox{\textwidth}{!}{%
% \begin{tabular}{c|cccccccccccc}
% \toprule
% Uncertainty & \cellcolor{lightgray} Lik. & \cellcolor{lightgray} Cal. & \cellcolor{lightgray} Lik. & \cellcolor{lightgray} Cal. & \cellcolor{lightgray} Lik. & \cellcolor{lightgray} Cal. & \cellcolor{lightgray} Lik. & \cellcolor{lightgray} Cal. & \cellcolor{lightgray} Lik. & \cellcolor{lightgray} Cal. \\
% \midrule 
% TFT & -- & \textcolor{red}{+94.6\%} & -- & \textcolor{red}{+114.61\%} & -- & \textcolor{red}{+7.57\%} & -- & \textcolor{red}{+16.84\%} & -- & \textcolor{blue}{-21.55\%}
% \\
% \bottomrule
% \end{tabular}
% }
% \end{table}
% \input{table_short_std}

\subsection{Why does the performance of the models differ between the datasets?} %(Nathaniel)
\label{subsec:model-failure}

Three factors consistently impact the results across the datasets and model configurations: (1)~dataset size, (2) patients'
 composition, and (3) time of day. Below we discuss the effects of these factors on accuracy (Task 1), similar observations hold for uncertainty quantification (Task~2).

{Tables~\ref{table:results-short-acc}}
% and \ref{table:results-short-uq}}
indicates that the best-performing model on each dataset is dependent on the dataset size. For smaller datasets, such as \citet{broll2021interpreting} and \citet{dubosson2018open}, simple models like ARIMA and linear regression yield the best results. {In general, we see that deep learning models excel on larger datasets: \citet{hall2018glucotypes} (best model is Latent ODE) and \citet{weinstock2016risk} (best model is Transformer) are 2 of the largest datasets. The only exception is \citet{colas2019detrended}, on which the best model is linear regression. We suggest that this could be explained by the fact that despite being large, \citet{colas2019detrended} dataset has low number observations per patient: 100,000 glucose readings across 200 patients yeilds 500 readings or 2 days worth of data per patient. In comparison, \citet{hall2018glucotypes} has 2,000 readings per patient or 7 days, and \citet{weinstock2016risk} has approximately 3,000 readings per patient or 10 days.}

Figure~\ref{fig:analysis}(b) demonstrates that the accuracy of predictions is substantially influenced by the patients' population group. Healthy subjects demonstrate markedly smaller errors compared to subjects with Type 1 or Type 2 diabetes. This discrepancy is due to healthy subjects maintaining a narrower range of relatively low glucose level, simplifying forecasting. Patients with Type 1 exhibit larger fluctuations partly due to consistently required insulin administration in addition to lifestyle-related factors, whereas most patients with Type 2 are not on insulin therapy.

Figure~\ref{fig:analysis}(c) shows the impact of time of day on accuracy, with daytime (defined as 9:00AM to 9:00PM) being compared to nighttime for Transformer model on \citet{broll2021interpreting} dataset. The distribution of daytime errors is more right-skewed and right-shifted compared to the distribution of nighttime errors, signifying that daytime glucose values are harder to predict. Intuitively, glucose values are less variable overnight due to the absence of food intake and exercise, simplifying forecasting.
We include similar plots for all models and datasets in Appendix~B. This finding underscores the importance of accounting for daytime and nighttime when partitioning CGM data for model training and evaluation.

Overall, we recommend using simpler shallow models when data is limited, the population group exhibits less complex CGM profiles (such as healthy individuals or Type 2 patients), or for nighttime forecasting. Conversely, when dealing with larger and more complex datasets, deep or hybrid models are the preferred choice. In clinical scenarios where data is actively collected, it is advisable to deploy simpler models during the initial stages and, in later stages, maintain an ensemble of both shallow and deep models. The former can act as a guardrail or be used for nighttime predictions.

 %Generally, biological signals display lower variability during nighttime, assuming the subject is resting. 
 
 % The inherent difference between daytime and nighttime in the difficulty of forecasting is behind lower OD errors for \citet{dubosson2018open} dataset compared to ID errors. We found that the testing splits for \citet{dubosson2018open} contained a significantly larger portion of nighttime data compared to the training splits. This finding underscores the importance of accounting for daytime and nighttime when partitioning CGM data for model training and evaluation.

\subsection{Are the models generalizable to patients beyond the training dataset?} %(Nathaniel)
\label{subsec:generalizability}

{Table~\ref{table:results-short-acc-od}}
% and \ref{table:results-short-uq-od}} 
compares accuracy and uncertainty quantification of selected models on in-distribution (ID) and out-of-distribution (OD) test sets, while the full table is provided in Appendix~B. 
% Performance increases are indicated in blue, while decreases in red. 
% We note that any potential positive changes could be attributed to two sources: (1) unmeasured similarity and composition of ID and OD sets, (2) model generalization power. 
Here we assume that each patient is different, in that the OD set represents a distinct distribution from the ID set. 

% Recall from Section~\ref{pre-processing} that if patient data was used for training, we call such set ID, otherwise all patient data that has been witheld from the model becomes an OD set. 
% {From Section~\ref{pre-processing}, we recall that if any part of the data for a patient was used for training, we call the corresponding test set ``in-distribution." If the data for a patient was completely withheld from the model at all stages (training / validation), then we call such test set ``out-of-distribution."}

In both tasks, most models exhibit decreased performance on the OD data, emphasizing individual-level variation between patients and the difficulty of cold starts on new patient populations. Figure~\ref{fig:analysis}(a) displays OD-to-ID accuracy ratio (measured in MAE) for each model and dataset: higher ratios indicate poorer generalization, while lower ratios indicate better generalization. In general, we observe that deep learning models (Transformer, NHiTS, TFT, Gluformer, and Latent ODE) generalize considerably better than the simple baselines (ARIMA and linear regression). We attribute this to the deep learning models' ability to capture and recall more patterns from the data. Notably, XGBoost also demonstrates strong generalization capabilities and, in some instances, outperforms the deep learning models in the generalization power.
%Interestingly, the Transformer model shows some resilience, with improved accuracy in certain cases. Additionally, we provide a plot of  in . From the plot,  

% On Task 2, Gluformer has improved likelihood on all datasets except \citet{colas2019detrended}, suggesting it is well-equipped to quantify uncertainty for OD data. On the other hand, the calibration results are mixed for both the Gluformer and TFT models. 

\subsection{How does adding the covariates affect the modeling quality?} %(Valeriya)
\label{subsec:covariate-effects}

% In here, the idea is to compare Table 1 (with no covariates) and Table 2 (with covariates). Things to keep in mind:
% \begin{enumerate}
%     \item Not all models can handle covariates (look up in DARTS which ones can)
%     \item Each dataset has different set of covariates: dynamic, static etc
% \end{enumerate}
% In general, with this subsection, we want to answer whether adding covariates helps and quantify that statement.

{Table~\ref{table:results-short-acc-cov}}
% and \ref{table:results-short-uq-cov}} 
demonstrates the impact of including covariates in the models on Task 1 (accuracy) and Task~2 (uncertainty quantification) compared to the same models with no covariates. As the inclusion of covariates represents providing model with more information, any changes in performance can be attributed to (1) the quality of the covariate data; (2) model's ability to handle multiple covariates. We omit ARIMA, Gluformer, and Latent ODE models as their implementations do not support covariates.

In both tasks, the impact of covariates on model performance varies depending on the dataset. For \citet{colas2019detrended} and \citet{dubosson2018open}, we observe a decrease in both accuracy and uncertainty quantification performance with the addition of covariates. Given that these are smaller datasets with a limited number of observations per patient, we suggest that the inclusion of covariates leads to model overfitting, consequently increasing test-time errors. In contrast, for \citet{broll2021interpreting} that is also small, unlike for all other datasets, we have covariates extracted solely from the timestamp, which appears to enhance model accuracy. 
This increase in performance is likely attributable to all patients within the train split exhibiting more pronounced cyclical CGM patterns, which could explain why the overfitted model performs better. This is further supported by the fact that the performance on the OD set deteriorates with the addition of covariates. Finally, in the case of \citet{hall2018glucotypes} and \citet{weinstock2016risk}, which are large datasets, the inclusion of covariates has mixed effects, indicating that covariates do not contribute significantly to the model's performance.

% While the Linear Regression and Transformer models exhibit improved performance on some datasets (\citet{broll2021interpreting} and \citet{weinstock2016risk} for Linear; \citet{broll2021interpreting} and \citet{hall2018glucotypes} for Transformer), they show decreased performance on others, suggesting that the inclusion of covariates has dataset-dependent effect on accuracy. On Task 2, the calibration performance of TFT worsens in \citet{broll2021interpreting}, \citet{colas2019detrended}, \citet{dubosson2018open}, and \citet{hall2018glucotypes} datasets, but improves in \citet{weinstock2016risk} dataset. This indicates that the effect of covariates on uncertainty quantification is also dataset-dependent. 

\section{Discussion}
\label{sec:discussion}
\textbf{Impact.} We discuss potential negative societal impact of our work. {\textbf{First}, inaccurate glucose forecasting could lead to severe consequences for patients. This is by far the most important consideration that we discuss further in Appendix D.}
% {Implementing predictive models in automatic systems, such as the Artificial Pancreas, without thorough checks or proper education on their limitations carries a potentially lethal risk. An error in the model could prompt the system to overcompensate with insulin, leading to serious health consequences. Similarly, the absence of adequate education about the model's limitations may result in individuals trusting mistaken predictions, potentially causing them to overcompensate with insulin and posing a significant health risk. Emphasizing the need for rigorous checks and comprehensive user education is crucial to mitigate these potentially life-threatening consequences.} 
{\textbf{Second}, there is a potential threat from CGM device hacking that could affect model predictions.}
% Therefore, it is crucial to exercise extreme caution when relying on glucose forecasting for diabetes management.}
\textbf{Third}, the existence of pre-defined tasks and datasets may stifle research, as researchers might focus on overfitting and marginally improving upon well-known datasets and tasks. {\textbf{Finally}, the release of health records must be treated with caution to guarantee patients' right to privacy.}
% Ensuring the preservation of patient anonymity and preventing de-identification are essential to protect the privacy and security of patients.
% {In addition, there is a potential danger of AP devices hardware or software systems malfunctioning or being infected with malware which could lead to insulin overdose with lethal outcome.}

\textbf{Future directions.} We outline several research avenues: (1) adding new public CGM datasets and tasks; (2) open-sourcing physiological and hybrid models; (3) exploring model training augmentation, such as pre-training on aggregated data followed by patient-specific fine-tuning and down-sampling night periods; (4) developing scaling laws for dataset size and model performance; and (5) examining covariate quality and principled integration within models. {Related to the point (5), we note that out of the 5 collected datasets, only \citet{dubosson2018open} records time-varying covariates describing patients physical activity (e.g. accelerometer readings, heart rate), blood pressure, food intake, and medication. We believe having larger datasets that comprehensively track dynamic patient behavior could lead to new insights and more accurate forecasting.}

\section{Conclusion} In this work, we have presented a comprehensive resource to address the challenges in CGM-based glucose trajectory prediction, including a curated repository of public datasets, a standardized task list, a set of benchmark models, and a detailed analysis of performance-influencing factors. Our analysis emphasizes the significance of dataset size, patient population, testing splits (e.g., in- and out-of-distribution, daytime, nighttime), and covariate availability. 

% We believe that this resource will serve as a valuable foundation for future research and practical adoption of effective glucose prediction models, ultimately benefiting individuals with diabetes.
%  By offering open-source access, we aim to foster reproducibility, accessibility, and collaboration within the research community. 

\section*{Acknowledgements}
The source of a subset of the data is the T1D Exchange, but the analyses, content, and conclusions presented herein are solely the responsibility of the authors and have not been reviewed or approved by the T1D Exchange.

% We thank Akhil Robertson Cutinha and Urjeet Shrestha for their help with the datasets processing, model prototyping, and organizing GitHub repository. This research was partially supported by NSF DMS-2044823.

\newpage
\appendix
\section{Datasets}
\label{appendix:a}

\textbf{Previous works.} We summarize previous work on the CGM datasets in Table~\ref{table:related-works}.

\begin{table}[h]
    \caption{Summary of the glucose prediction models by dataset and model type. %For models, we denote "deep" for deep learning, "shallow" for other non-deep learning, and "physiological" for physiological models.
    We indicate "open" for datasets that are publicly available online, and "proprietary" for the ones that cannot be released.}
    \label{table:related-works}
    \centering
            \small
    \begin{tabular}{ p{2cm} | p{0.9cm} p{0.8cm} | p{2.5cm} p{2.5cm} p{2.5cm}}
        \toprule
        \multicolumn{1}{c|}{Dataset} & \multicolumn{1}{c}{Diabetes} & \multicolumn{1}{c|}{ \#} & \multicolumn{1}{c}{Deep} & \multicolumn{1}{c}{Shallow} & \multicolumn{1}{c}{Physiological} \\ 

        \midrule
        %DirecNet 
        \citetalias{diabetes2008use} & Type 1 & 30 
        & \citet{heBloodGlucoseConcentration2020}
        & \citet{eren-orukluHypoglycemiaPredictionSubjectSpecific2010}
        & %\RaggedRight{
        \citet{balakrishnanPersonalizedHybridModels2013, chen2010modeling}
        %}
        \\
        \cmidrule(lr){4-4} \cmidrule(lr){5-5} \cmidrule(lr){6-6}
        %Anthimopoulos 
        \citet{anthimopoulos2015computer} & Type 1 &  20
        & \citet{sunPredictingBloodGlucose2018}
        & 
        & 
        \\
        \cmidrule(lr){4-4} \cmidrule(lr){5-5} \cmidrule(lr){6-6}
        %Mauras 
        \citet{mauras2012randomized} & Type 1 & 146 
        & \citet{indrawanBloodGlucosePrediction2021} 
        &
        &
        \\
        \cmidrule(lr){4-4} \cmidrule(lr){5-5} \cmidrule(lr){6-6}
        %Protopappas 
        \citet{georga2009data} & Type 1 & 15 
        & 
        & \citet{georgaEvaluationShorttermPredictors2015}
        & 
        \\
        \cmidrule(lr){4-4} \cmidrule(lr){5-5} \cmidrule(lr){6-6}
        %Marling 
        \citet{marling2020ohiot1dm} & Type 1 & 12 
        & %\RaggedRight{
        \citet{dengDeepTransferLearning2021, vandoornMachineLearningbasedGlucose2021, zhuDeepLearningAlgorithma, martinssonBloodGlucosePrediction2020}
        %}
        & \citet{vandoornMachineLearningbasedGlucose2021}
        & 
        \\
        \cmidrule(lr){4-4} \cmidrule(lr){5-5} \cmidrule(lr){6-6}
        %Dubosson 
        \citet{dubosson2018open} & Type 1 & 9  
        & \citet{munoz-organeroDeepPhysiologicalModel2020} 
        & 
        & 
        \\
        \cmidrule(lr){4-4} \cmidrule(lr){5-5} \cmidrule(lr){6-6}
        %Aleppo 
        \citet{aleppo2017replace} & Type 1 & 168
        & \citet{jaloliLongtermPredictionBlood2022}
        & 
        & 
        \\
        \cmidrule(lr){4-4} \cmidrule(lr){5-5} \cmidrule(lr){6-6}
        %Fox 
        \citet{foxDeepMultiOutputForecasting2018a} & Type 1 & 40 
        & %\RaggedRight{
        \citet{foxDeepMultiOutputForecasting2018a,armandpourDeepPersonalizedGlucose2021a, sergazinov2022gluformer}
        %}
        & 
        & 
        \\
        \cmidrule(lr){4-4} \cmidrule(lr){5-5} \cmidrule(lr){6-6}
        %Cescon 
        \citet{cescon2013modeling} & Type 1 & 59 
        & \citet{jaloliLongtermPredictionBlood2022}
        & 
        & 
        \\
        \midrule
        \rowcolor{lightgray}
        Total (open) & & & 
        \multicolumn{1}{c}{\textcolor{blue}{13}} & 
        \multicolumn{1}{c}{\textcolor{blue}{3}} & 
        \multicolumn{1}{c}{\textcolor{blue}{2}} \\
        \midrule 
        Simulation & NA & NA 
        & %\RaggedRight{
        \citet{liConvolutionalRecurrentNeural2020, langaricaMetalearningApproachPersonalized2023, liuEnhancingBloodGlucose2018}
        %}
        & %\RaggedRight{
        \citet{reymannBloodGlucoseLevel2016, boirouxOvernightControlBlood2012, otoomRealTimeStatisticalModeling2015}
        %}
        & %\RaggedRight{
        \citet{boirouxOvernightControlBlood2012, bockTherapyParameterbasedModel2015, calm2011comparison, de2012prediction, fang2015new, laguna2014identification}
        %}
        \\
        \cmidrule(lr){4-4} \cmidrule(lr){5-5} \cmidrule(lr){6-6}
        Proprietary & NA
        & 1-851 
        % 4, 5, 10, 12, 26, 33, 40, 100, 451, 851
        & %\RaggedRight{
        \citet{xuBloodGlucosePrediction2022, liConvolutionalRecurrentNeural2020, prendinForecastingGlucoseLevels2021, alibertiMultiPatientDataDrivenApproach2019, liuEnhancingBloodGlucose2018, benaliContinuousBloodGlucose2018, shi2015glucose}
        %}
        & %\RaggedRight{
        \citet{prendinForecastingGlucoseLevels2021, yangARIMAModelAdaptive2019, sudharsanHypoglycemiaPredictionUsing2015, hidalgoDataBasedPrediction2017, efendic2014short, botwey2014multi, wang2013novel, zarkogianni2014neuro}
        %} 
        & %\RaggedRight{
        \citet{gyukBloodGlucoseLevel2019, novaraNonlinearBlindIdentification2016, bockTherapyParameterbasedModel2015, duun2013model, laguna2014experimental, wu2011physiological, zhang2016data}
        %}
        \\
        \midrule
        \rowcolor{lightgray}
        Total (proprietary) & & & 
        \multicolumn{1}{c}{\textcolor{blue}{10}} & 
        \multicolumn{1}{c}{\textcolor{blue}{11}} & 
        \multicolumn{1}{c}{\textcolor{blue}{13}} \\
        \bottomrule
    \end{tabular}
\end{table}

\textbf{Access.} The datasets are distributed according to the following licences and can be downloaded from the following links:

\begin{tabular}{l l l l}
1. & \citet{broll2021interpreting}      & License: \href{https://www.r-project.org/Licenses/GPL-2}{GPL-2} & Source: \href{https://github.com/irinagain/iglu}{link} \\
2. & \citet{colas2019detrended}     & License: \href{https://creativecommons.org/licenses/by/3.0/us/}{Creative Commons 4.0} & Source: \href{https://journals.plos.org/plosone/article?id=10.1371/journal.pone.0225817#sec018}{link} \\
3. & \citet{dubosson2018open}  & License: \href{https://creativecommons.org/licenses/by-sa/4.0/legalcode}{Creative Commons 4.0} & Source: \href{https://doi.org/10.5281/zenodo.1421615}{link} \\
4. &   \citet{hall2018glucotypes}    & License: \href{https://creativecommons.org/licenses/by/4.0/}{Creative Commons 4.0} & Source: \href{https://journals.plos.org/plosbiology/article?id=10.1371/journal.pbio.2005143#pbio.2005143.s010}{link} \\
5. & \citet{weinstock2016risk}  & License: \href{https://creativecommons.org/licenses/by-sa/4.0/legalcode}{Creative Commons 4.0} & Source: \href{https://public.jaeb.org/dataset/537}{link} \\
\end{tabular}

\textbf{Covariates.} We summarize covariate types for each dataset in Table~\ref{table:covariate-info}. For each dataset, we extract the following dynamic known covariates from the time stamp: year, month, day, hour, minute, and second (only for Broll). Broll provides no covariates aside from the ones extracted from the time stamp. Colas, Hall, and Weinstock provide demographic information for the patients (static covariates). Dubosson is the only dataset for which dynamic unknown covariates such as heart rate, insulin levels, and blood pressure are available. 

\begin{table}[!h]
    \caption{Covariate information for each dataset.}
    \label{table:covariate-info}
    \centering
    \begin{tabular}{ c c | c c c c c}
    \toprule
    & Covariate &  Broll & Colas & Dubosson & Hall & Weinstock\\
    \midrule
    \multirow{3}{*}{\rotatebox{90}{Static}} & Age & & \checkmark & & \checkmark \\
    & Height & & & & \checkmark & \checkmark \\
    & ... & \\
    \midrule
    \rowcolor{lightgray}
    \multicolumn{2}{c|}{Total} & 0 & 7 & 0 & 48 & 38 \\
    \midrule
    \multirow{4}{*}{\rotatebox{90}{\parbox{1em}{Dyn. Kn.}}}
    & Year & \checkmark & \checkmark & \checkmark & \checkmark & \checkmark \\
    & Month & \checkmark & \checkmark & \checkmark & \checkmark & \checkmark \\
    & ... & \\
    \midrule
    \rowcolor{lightgray}
    \multicolumn{2}{c|}{Total} & 6 & 5 & 5 & 5 & 5\\
    \midrule
    \multirow{4}{*}{\rotatebox{90}{\parbox{1em}{Dyn. Unkn.}}} & Insulin & & & \checkmark & & \\
    & Heart Rate & & & \checkmark & & \\
    & ... & \\
    \midrule
    \rowcolor{lightgray}
    \multicolumn{2}{c|}{Total} & 0 & 0 & 11 & 0 & 0 \\
    \bottomrule
    \end{tabular}
\end{table}

\clearpage

\section{Analysis}
\label{appendix:b}

\subsection{Visualized predictions}
{
We provide visualized forecasts for the same 5 segments of \citet{weinstock2016risk} data for the best performing models: linear regression, Latent ODE \citep{rubanova2019latent}, and Transformer \citep{vaswani2017attention} on Task 1 (accuracy), and Gluformer \citep{sergazinov2022gluformer} and TFT \citep{lim2021temporal} on Task 2 (uncertainty). For the best models on Task 2, we also provide the estimated confidence intervals or the predictive distribution, whichever is available. For visualization, we have truncated the input sequence to 1 hour (12 points); however, we note that different models have different input length and require at least 4 hours of observations to forecast the future trajectory.}

\begin{figure}[h]
    \centering
    \includegraphics[width=\textwidth]{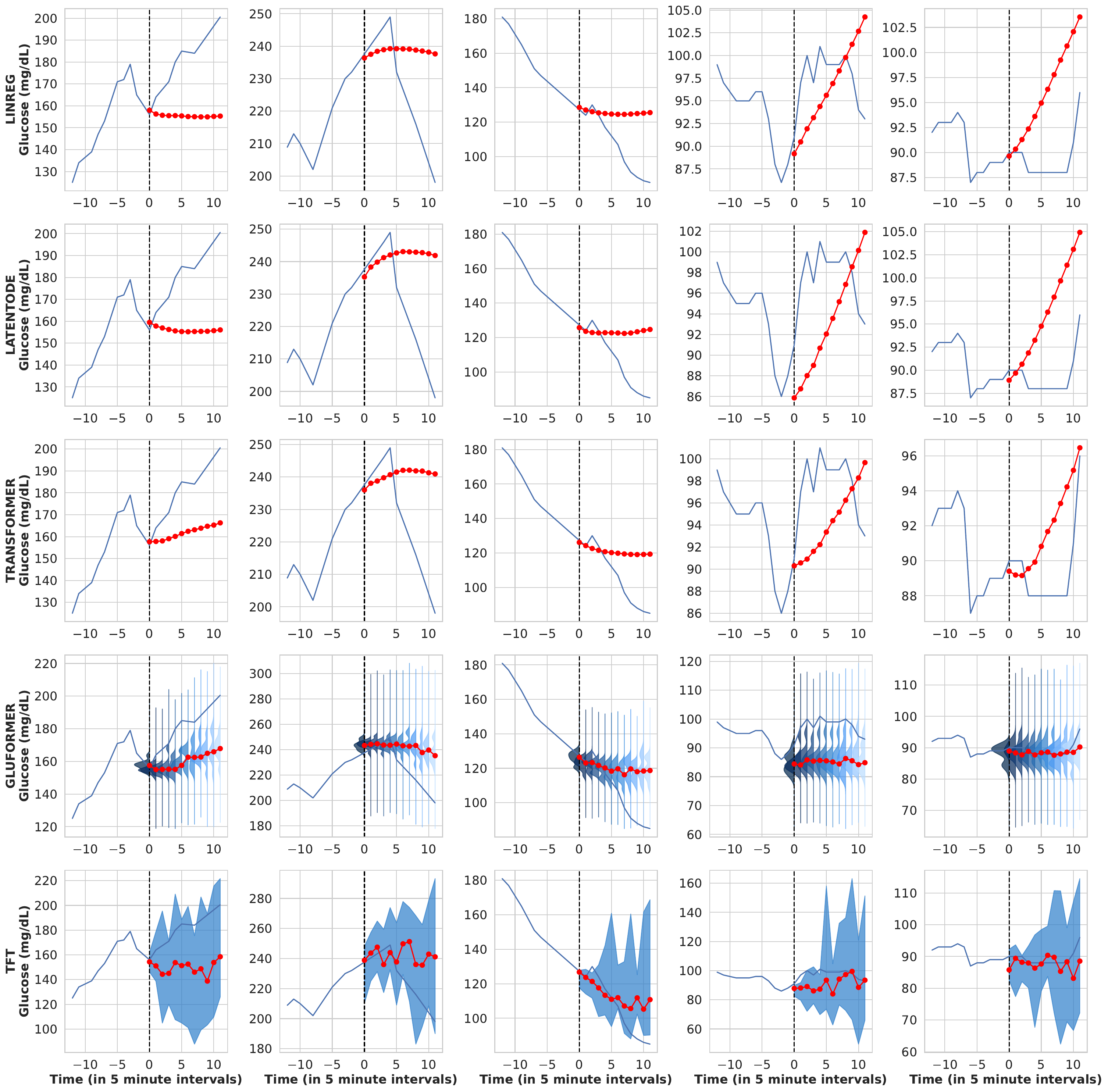}
    \caption{Model forecasts on \citet{weinstock2016risk} dataset.}
    \label{fig:enter-label}
\end{figure}

\subsection{Performance}
For reference, we include results for all models both with and without covariates evaluated on ID and OD splits in Table~\ref{table:results-accuracy} on Task 1 (accuracy) and in Table~\ref{table:results-likelihood} on Task 2 (uncertainty quantification). 
% We provide the following guidelines for reading the tables: (1) to compare the effect of adding covariates, locate the row denoted by "Improv.", (2) row $\min\Delta$(ID, OD)\% indicates the best generalization error from ID to OD among the pair of models with and without covariates, (3) the increase in performance is indicated by \textcolor{blue}{blue} and decrease by \textcolor{red}{red}.

\begin{table}[h]
{
\caption{Model results on the data sets for Task 1 (accuracy).}
% , $p(\cdot|x)$ denotes inclusion of covariates such as demographic information. To compare models with and without covariates, we include the percent improvement on ID and OD test splits. To compare the performance on the ID versus on the OD data, we compute the smallest percent difference between ID and OD for models with and without covariates, denoted as $\min \Delta$(ID, OD)\%
\label{table:results-accuracy}
\centering
\resizebox{\textwidth}{!}{%
\begin{tabular}{ccc|cccccccccccc}
\toprule
 & \multirow{2}{*}{$p(\cdot | x)$} & \multirow{2}{*}{Data} & \multicolumn{2}{c}{Broll} & \multicolumn{2}{c}{Colas} & \multicolumn{2}{c}{Dubosson} & \multicolumn{2}{c}{Hall} & \multicolumn{2}{c}{Weinstock}  \\ 
\cmidrule(lr){4-5} \cmidrule(lr){6-7} \cmidrule(lr){8-9} \cmidrule(lr){10-11} \cmidrule(lr){12-13}
& & & RMSE & MAE & RMSE & MAE & RMSE & MAE & RMSE & MAE & RMSE & MAE \\
\midrule

%%%%%%%%%%%%%%%%%%%%%%%%%%%%%%%%%%%%%%%%%%%%%%%%%%%%%%%%%%%%%%%%%%%%%
%%%%%%%%%%%%%%%%%%%%%%%%%%%%%%%%%%%%%%%%%%%%%%%%%%%%%%%%%%%%%%%%%%%%%
%%%%%%%%%%%%%%%%%%%%%%%%%%%%%%%%%%%%%%%%%%%%%%%%%%%%%%%%%%%%%%%%%%%%%
\multirow{2}{*}{\rotatebox{90}{ARI}} & \crossmark & 
ID & 10.53&8.67&5.80&4.80&13.53&11.06&8.63&7.34&13.40&11.25
\\
& \crossmark & 
OD & 11.75&9.71&5.91&4.87&18.75&14.58&8.22&6.97&15.87&13.34
\\
\midrule
\rowcolor{lightgray}
\multicolumn{3}{c|}{$\min \Delta$(ID, OD)\%}&
\multicolumn{2}{c}{\textcolor{red}{+11.8\%}} & \multicolumn{2}{c}{\textcolor{red}{+1.75\%}} & \multicolumn{2}{c}{\textcolor{red}{+35.18\%}} &  \multicolumn{2}{c}{\textcolor{blue}{-4.94\%}} & \multicolumn{2}{c}{\textcolor{red}{+18.51\%}}
\\
\midrule
\multirow{6}{*}{\rotatebox{90}{LIN}} & \crossmark & 
ID & 11.68&9.71&5.26&4.35&12.07&9.97&7.38&6.33&13.60&11.46
\\
& \checkmark & 
ID & 9.95&8.41&5.56&4.60&12.41&10.03&7.84&6.66&13.39&11.34
\\
\cmidrule(lr){4-5} \cmidrule(lr){6-7} \cmidrule(lr){8-9} \cmidrule(lr){10-11} \cmidrule(lr){12-13}
& \multicolumn{2}{c|}{Improv.} &
 \multicolumn{2}{c}{\textcolor{blue}{-14.08\%}} & \multicolumn{2}{c}{\textcolor{red}{+5.65\%}} & \multicolumn{2}{c}{\textcolor{red}{+1.73\%}} & \multicolumn{2}{c}{\textcolor{red}{+5.63\%}} &  \multicolumn{2}{c}{\textcolor{blue}{-1.31\%}}
\\
\cmidrule(lr){4-5} \cmidrule(lr){6-7} \cmidrule(lr){8-9} \cmidrule(lr){10-11} \cmidrule(lr){12-13}
& \crossmark &
OD & 11.98&9.83&5.33&4.41&15.69&11.90&7.86&6.62&15.58&13.09
\\
& \checkmark & 
OD & 23.30&16.80&5.54&4.57&203114.47&67548.59&14.22&10.02&15.66&13.16
\\
\cmidrule(lr){4-5} \cmidrule(lr){6-7} \cmidrule(lr){8-9} \cmidrule(lr){10-11} \cmidrule(lr){12-13}
& \multicolumn{2}{c|}{Improv.} &
\multicolumn{2}{c}{\textcolor{red}{+82.7\%}} & \multicolumn{2}{c}{\textcolor{red}{+3.8\%}} & \multicolumn{2}{c}{\textcolor{red}{+930929.99\%}} & \multicolumn{2}{c}{\textcolor{red}{+66.07\%}} & \multicolumn{2}{c}{\textcolor{red}{+0.54\%}}
\\
\midrule
\rowcolor{lightgray}
\multicolumn{3}{c|}{$\min \Delta$(ID, OD)\%}&
\multicolumn{2}{c}{\textcolor{red}{+1.9\%}} &  \multicolumn{2}{c}{\textcolor{blue}{-0.45\%}} & \multicolumn{2}{c}{\textcolor{red}{+24.71\%}} & \multicolumn{2}{c}{\textcolor{red}{+5.52\%}} & \multicolumn{2}{c}{\textcolor{red}{+14.36\%}}
\\
\midrule
\multirow{6}{*}{\rotatebox{90}{XGB}} & \crossmark & 
ID & 12.80&11.50&6.42&5.49&21.18&19.09&7.58&6.55&13.63&11.61
\\
& \checkmark & 
ID & 13.89&11.87&6.37&5.46&20.89&18.55&8.05&7.02&13.77&11.77
\\
\cmidrule(lr){4-5} \cmidrule(lr){6-7} \cmidrule(lr){8-9} \cmidrule(lr){10-11} \cmidrule(lr){12-13}
& \multicolumn{2}{c|}{Improv.} &
\multicolumn{2}{c}{\textcolor{red}{+5.88\%}} &  \multicolumn{2}{c}{\textcolor{blue}{-0.61\%}} &  \multicolumn{2}{c}{\textcolor{blue}{-2.08\%}} & \multicolumn{2}{c}{\textcolor{red}{+6.67\%}} & \multicolumn{2}{c}{\textcolor{red}{+1.18\%}}
\\
\cmidrule(lr){4-5} \cmidrule(lr){6-7} \cmidrule(lr){8-9} \cmidrule(lr){10-11} \cmidrule(lr){12-13}
& \crossmark &
OD & 9.76&8.72&6.18&5.32&17.57&15.42&7.49&6.52&15.36&13.04
\\
& \checkmark & 
OD & 9.67&8.56&6.36&5.47&17.44&15.46&8.20&7.11&15.55&13.43
\\
\cmidrule(lr){4-5} \cmidrule(lr){6-7} \cmidrule(lr){8-9} \cmidrule(lr){10-11} \cmidrule(lr){12-13}
& \multicolumn{2}{c|}{Improv.} &
 \multicolumn{2}{c}{\textcolor{blue}{-1.37\%}} & \multicolumn{2}{c}{\textcolor{red}{+2.91\%}} &  \multicolumn{2}{c}{\textcolor{blue}{-0.26\%}} & \multicolumn{2}{c}{\textcolor{red}{+9.25\%}} & \multicolumn{2}{c}{\textcolor{red}{+2.09\%}}
\\
\midrule
\rowcolor{lightgray}
\multicolumn{3}{c|}{$\min \Delta$(ID, OD)\%}&
 \multicolumn{2}{c}{\textcolor{blue}{-29.14\%}} &  \multicolumn{2}{c}{\textcolor{blue}{-3.47\%}} &  \multicolumn{2}{c}{\textcolor{blue}{-18.11\%}} &  \multicolumn{2}{c}{\textcolor{blue}{-0.84\%}} & \multicolumn{2}{c}{\textcolor{red}{+12.51\%}}
\\
\midrule
\multirow{2}{*}{\rotatebox{90}{GLU}} & \crossmark & 
ID & 14.19&12.55&8.17&7.12&21.74&19.40&7.74&6.69&14.07&12.09
\\
& \crossmark & 
OOD & 16.70&14.82&6.94&6.03&23.48&20.70&8.17&7.04&15.94&13.65
\\
\midrule
\rowcolor{lightgray}
\multicolumn{3}{c|}{$\min \Delta$(ID, OD)\%}&
\multicolumn{2}{c}{\textcolor{red}{+17.85\%}} &  \multicolumn{2}{c}{\textcolor{blue}{-15.17\%}} & \multicolumn{2}{c}{\textcolor{red}{+7.37\%}} & \multicolumn{2}{c}{\textcolor{red}{+5.39\%}} & \multicolumn{2}{c}{\textcolor{red}{+13.08\%}}
\\
\midrule
\multirow{2}{*}{\rotatebox{90}{LAT}} & \crossmark & 
ID & 14.37&12.32&6.28&5.37&20.14&17.88&7.13&6.11&13.54&11.45
\\
& \crossmark & 
OOD & 14.96&13.05&5.64&4.84&17.38&15.12&7.71&6.61&15.06&12.72
\\
\midrule
\rowcolor{lightgray}
\multicolumn{3}{c|}{$\min \Delta$(ID, OD)\%}&
\multicolumn{2}{c}{\textcolor{red}{+5.03\%}} &  \multicolumn{2}{c}{\textcolor{blue}{-9.94\%}} &  \multicolumn{2}{c}{\textcolor{blue}{-14.59\%}} & \multicolumn{2}{c}{\textcolor{red}{+8.12\%}} & \multicolumn{2}{c}{\textcolor{red}{+11.12\%}}
\\
\midrule
\multirow{6}{*}{\rotatebox{90}{NHI}} & \crossmark & 
ID & 13.79&12.07&5.93&5.04&17.45&14.79&7.68&6.57&13.29&11.21
\\
& \checkmark & 
ID & 16.20&14.64&9.09&8.03&30.43&27.97&8.16&7.10&13.41&11.31
\\
\cmidrule(lr){4-5} \cmidrule(lr){6-7} \cmidrule(lr){8-9} \cmidrule(lr){10-11} \cmidrule(lr){12-13}
& \multicolumn{2}{c|}{Improv.} &
\multicolumn{2}{c}{\textcolor{red}{+19.36\%}} & \multicolumn{2}{c}{\textcolor{red}{+56.31\%}} & \multicolumn{2}{c}{\textcolor{red}{+81.74\%}} & \multicolumn{2}{c}{\textcolor{red}{+7.13\%}} & \multicolumn{2}{c}{\textcolor{red}{+0.9\%}}
\\
\cmidrule(lr){4-5} \cmidrule(lr){6-7} \cmidrule(lr){8-9} \cmidrule(lr){10-11} \cmidrule(lr){12-13}
& \crossmark &
OD & 14.64&12.77&5.68&4.83&18.20&15.59&7.74&6.62&14.52&12.24
\\
& \checkmark & 
OD & 15.66&14.01&7.56&6.65&37.35&33.52&8.59&7.53&14.40&12.12
\\
\cmidrule(lr){4-5} \cmidrule(lr){6-7} \cmidrule(lr){8-9} \cmidrule(lr){10-11} \cmidrule(lr){12-13}
& \multicolumn{2}{c|}{Improv.} &
\multicolumn{2}{c}{\textcolor{red}{+8.35\%}} & \multicolumn{2}{c}{\textcolor{red}{+35.49\%}} & \multicolumn{2}{c}{\textcolor{red}{+110.09\%}} & \multicolumn{2}{c}{\textcolor{red}{+12.34\%}} &  \multicolumn{2}{c}{\textcolor{blue}{-0.91\%}}
\\
\midrule
\rowcolor{lightgray}
\multicolumn{3}{c|}{$\min \Delta$(ID, OD)\%}&
 \multicolumn{2}{c}{\textcolor{blue}{-3.8\%}} &  \multicolumn{2}{c}{\textcolor{blue}{-17.01\%}} & \multicolumn{2}{c}{\textcolor{red}{+4.86\%}} & \multicolumn{2}{c}{\textcolor{red}{+0.78\%}} & \multicolumn{2}{c}{\textcolor{red}{+7.29\%}}
\\
\midrule
\multirow{6}{*}{\rotatebox{90}{TFT}} & \crossmark & 
ID & 13.73&11.07&5.62&4.54&18.37&15.49&7.92&6.61&14.32&11.76
\\
& \checkmark & 
ID & 14.68&12.43&6.51&5.27&18.43&15.51&8.42&7.06&14.97&12.30
\\
\cmidrule(lr){4-5} \cmidrule(lr){6-7} \cmidrule(lr){8-9} \cmidrule(lr){10-11} \cmidrule(lr){12-13}
& \multicolumn{2}{c|}{Improv.} &
\multicolumn{2}{c}{\textcolor{red}{+9.53\%}} & \multicolumn{2}{c}{\textcolor{red}{+15.9\%}} & \multicolumn{2}{c}{\textcolor{red}{+0.22\%}} & \multicolumn{2}{c}{\textcolor{red}{+6.52\%}} & \multicolumn{2}{c}{\textcolor{red}{+4.55\%}}
\\
\cmidrule(lr){4-5} \cmidrule(lr){6-7} \cmidrule(lr){8-9} \cmidrule(lr){10-11} \cmidrule(lr){12-13}
& \crossmark &
OD & 12.43&10.23&5.51&4.47&17.50&14.53&8.12&6.76&15.25&12.50
\\
& \checkmark & 
OD & 13.25&11.17&5.79&4.68&17.19&14.43&8.93&7.44&15.47&12.67
\\
\cmidrule(lr){4-5} \cmidrule(lr){6-7} \cmidrule(lr){8-9} \cmidrule(lr){10-11} \cmidrule(lr){12-13}
& \multicolumn{2}{c|}{Improv.} &
\multicolumn{2}{c}{\textcolor{red}{+7.91\%}} & \multicolumn{2}{c}{\textcolor{red}{+4.84\%}} &  \multicolumn{2}{c}{\textcolor{blue}{-1.22\%}} & \multicolumn{2}{c}{\textcolor{red}{+10.01\%}} & \multicolumn{2}{c}{\textcolor{red}{+1.41\%}}
\\
\midrule
\rowcolor{lightgray}
\multicolumn{3}{c|}{$\min \Delta$(ID, OD)\%}&
 \multicolumn{2}{c}{\textcolor{blue}{-9.91\%}} &  \multicolumn{2}{c}{\textcolor{blue}{-11.11\%}} &  \multicolumn{2}{c}{\textcolor{blue}{-6.84\%}} & \multicolumn{2}{c}{\textcolor{red}{+2.41\%}} & \multicolumn{2}{c}{\textcolor{red}{+3.18\%}}
\\
\midrule
\multirow{6}{*}{\rotatebox{90}{TRA}} & \crossmark & 
ID & 15.12&13.20&6.47&5.65&16.62&14.04&7.89&6.78&13.22&11.22
\\
& \checkmark & 
ID & 12.83&11.27&8.44&7.77&27.43&24.40&7.49&6.42&14.46&12.61
\\
\cmidrule(lr){4-5} \cmidrule(lr){6-7} \cmidrule(lr){8-9} \cmidrule(lr){10-11} \cmidrule(lr){12-13}
& \multicolumn{2}{c|}{Improv.} &
 \multicolumn{2}{c}{\textcolor{blue}{-14.89\%}} & \multicolumn{2}{c}{\textcolor{red}{+33.93\%}} & \multicolumn{2}{c}{\textcolor{red}{+69.41\%}} &  \multicolumn{2}{c}{\textcolor{blue}{-5.23\%}} & \multicolumn{2}{c}{\textcolor{red}{+10.87\%}}
\\
\cmidrule(lr){4-5} \cmidrule(lr){6-7} \cmidrule(lr){8-9} \cmidrule(lr){10-11} \cmidrule(lr){12-13}
& \crossmark &
OD & 14.04&12.28&5.97&5.24&15.71&12.98&8.18&7.07&14.15&11.91
\\
& \checkmark & 
OD & 13.76&12.13&7.26&6.59&34.11&28.21&7.40&6.29&15.59&13.58
\\
\cmidrule(lr){4-5} \cmidrule(lr){6-7} \cmidrule(lr){8-9} \cmidrule(lr){10-11} \cmidrule(lr){12-13}
& \multicolumn{2}{c|}{Improv.} &
 \multicolumn{2}{c}{\textcolor{blue}{-1.59\%}} & \multicolumn{2}{c}{\textcolor{red}{+23.61\%}} & \multicolumn{2}{c}{\textcolor{red}{+117.27\%}} &  \multicolumn{2}{c}{\textcolor{blue}{-10.23\%}} & \multicolumn{2}{c}{\textcolor{red}{+12.14\%}}
\\
\midrule
\rowcolor{lightgray}
\multicolumn{3}{c|}{$\min \Delta$(ID, OD)\%}&
 \multicolumn{2}{c}{\textcolor{blue}{-7.06\%}} &  \multicolumn{2}{c}{\textcolor{blue}{-14.62\%}} &  \multicolumn{2}{c}{\textcolor{blue}{-6.52\%}} &  \multicolumn{2}{c}{\textcolor{blue}{-1.57\%}} & \multicolumn{2}{c}{\textcolor{red}{+6.58\%}}
\\

%%%%%%%%%%%%%%%%%%%%%%%%%%%%%%%%%%%%%%%%%%%%%%%%%%%%%%%%%%%%%%%%%%%%%
%%%%%%%%%%%%%%%%%%%%%%%%%%%%%%%%%%%%%%%%%%%%%%%%%%%%%%%%%%%%%%%%%%%%%
%%%%%%%%%%%%%%%%%%%%%%%%%%%%%%%%%%%%%%%%%%%%%%%%%%%%%%%%%%%%%%%%%%%%%
\bottomrule
  \end{tabular}}
}
\end{table}

\begin{table}[h]
\caption{Model results on the data sets for Task 2 (uncertainty quantification).}
% , $p(\cdot|x)$ denotes inclusion of covariates such as demographic information. To compare models with and without covariates, we include the percent improvement on ID and OD test splits. To compare the performance on the ID versus on the OD data, we compute the smallest percent difference between ID and OD for models with and without covariates, denoted as $\min \Delta$(ID, OD)\%
\label{table:results-likelihood}
\centering
\resizebox{\textwidth}{!}{%
\begin{tabular}{ccc|cccccccccc}
\toprule
 & \multirow{2}{*}{$p(\cdot | x)$} & \multirow{2}{*}{Data} & \multicolumn{2}{c}{Broll} & \multicolumn{2}{c}{Colas} & \multicolumn{2}{c}{Dubosson} & \multicolumn{2}{c}{Hall} & \multicolumn{2}{c}{Weinstock}  \\ 
\cmidrule(lr){4-5} \cmidrule(lr){6-7} \cmidrule(lr){8-9} \cmidrule(lr){10-11} \cmidrule(lr){12-13}
& & & Lik.$\uparrow$ & Cal.$\downarrow$ & Lik.$\uparrow$ & Cal.$\downarrow$ & Lik.$\uparrow$ & Cal.$\downarrow$ & Lik.$\uparrow$ & Cal.$\downarrow$ & Lik.$\uparrow$ & Cal.$\downarrow$ \\
\midrule

%%%%%%%%%%%%%%%%%%%%%%%%%%%%%%%%%%%%%%%%%%%%%%%%%%%%%%%%%%%%%%%%%%%%%
%%%%%%%%%%%%%%%%%%%%%%%%%%%%%%%%%%%%%%%%%%%%%%%%%%%%%%%%%%%%%%%%%%%%%
%%%%%%%%%%%%%%%%%%%%%%%%%%%%%%%%%%%%%%%%%%%%%%%%%%%%%%%%%%%%%%%%%%%%%
\multirow{2}{*}{\rotatebox{90}{ARI}} & \crossmark & 
ID & -9.93&0.11&-9.30&0.10&-10.47&0.10&-9.81&0.10&-10.21&0.12
\\
& \crossmark & 
OOD & -10.06&0.07&-9.38&0.08&-10.44&0.08&-9.66&0.07&-10.32&0.12
\\
\midrule
\rowcolor{lightgray}
\multicolumn{3}{c|}{$\min \Delta$(ID, OD)\%}&
\textcolor{red}{-1.30\%} & \textcolor{blue}{-36.06\%} & \textcolor{red}{-0.86\%} & \textcolor{blue}{-19.95\%} & \textcolor{blue}{+0.28\%} & \textcolor{blue}{-20.07\%} & \textcolor{blue}{+1.52\%} & \textcolor{blue}{-29.92\%} & \textcolor{red}{-1.07\%} & \textcolor{red}{+0.06\%}
\\
\midrule
\multirow{6}{*}{\rotatebox{90}{LIN}} & \crossmark & 
ID & -9.89&0.12&-9.19&0.15&-10.10&0.18&-9.56&0.10&-10.14&0.11
\\
& \checkmark & 
ID & -9.87&0.13&-9.17&0.19&-10.15&0.21&-10.30&0.19&-10.12&0.11
\\
\cmidrule(lr){4-5} \cmidrule(lr){6-7} \cmidrule(lr){8-9} \cmidrule(lr){10-11} \cmidrule(lr){12-13}
& \multicolumn{2}{c|}{Improv.} &
\textcolor{blue}{+0.15\%} & \textcolor{red}{+5.99\%} & \textcolor{blue}{+0.25\%} & \textcolor{red}{+24.47\%} & \textcolor{red}{-0.41\%} & \textcolor{red}{+11.39\%} & \textcolor{red}{-7.67\%} & \textcolor{red}{+97.36\%} & \textcolor{blue}{+0.13\%} & \textcolor{blue}{-1.67\%}
\\
\cmidrule(lr){4-5} \cmidrule(lr){6-7} \cmidrule(lr){8-9} \cmidrule(lr){10-11} \cmidrule(lr){12-13}
& \crossmark &
OD & -9.95&0.15&-9.16&0.15&-10.11&0.17&-9.53&0.10&-10.22&0.11
\\
& \checkmark & 
OD & -10.24&0.55&-9.16&0.17&-12.08&0.48&-10.42&0.23&-11.13&0.21
\\
\cmidrule(lr){4-5} \cmidrule(lr){6-7} \cmidrule(lr){8-9} \cmidrule(lr){10-11} \cmidrule(lr){12-13}
& \multicolumn{2}{c|}{Improv.} &
\textcolor{red}{-2.88\%} & \textcolor{red}{+256.79\%} & \textcolor{blue}{+0.01\%} & \textcolor{red}{+10.19\%} & \textcolor{red}{-19.52\%} & \textcolor{red}{+181.06\%} & \textcolor{red}{-9.26\%} & \textcolor{red}{+130.96\%} & \textcolor{red}{-8.92\%} & \textcolor{red}{+87.0\%}
\\
\midrule
\rowcolor{lightgray}
\multicolumn{3}{c|}{$\min \Delta$(ID, OD)\%}&
\textcolor{red}{-0.65\%} & \textcolor{red}{+24.98\%} & \textcolor{blue}{+0.33\%} & \textcolor{blue}{-8.88\%} & \textcolor{red}{-0.02\%} & \textcolor{blue}{-7.62\%} & \textcolor{blue}{+0.33\%} & \textcolor{red}{+4.43\%} & \textcolor{red}{-0.85\%} & \textcolor{blue}{-3.1\%}
\\
\midrule
\multirow{6}{*}{\rotatebox{90}{XGB}} & \crossmark & 
ID & -9.94&0.07&-9.42&0.10&-10.55&0.07&-9.68&0.09&-10.20&0.11
\\
& \checkmark & 
ID & -10.06&0.07&-9.40&0.09&-10.54&0.06&-9.70&0.09&-10.21&0.10
\\
\cmidrule(lr){4-5} \cmidrule(lr){6-7} \cmidrule(lr){8-9} \cmidrule(lr){10-11} \cmidrule(lr){12-13}
& \multicolumn{2}{c|}{Improv.} &
\textcolor{red}{-1.22\%} & \textcolor{red}{+0.59\%} & \textcolor{blue}{+0.12\%} & \textcolor{blue}{-7.2\%} & \textcolor{blue}{+0.13\%} & \textcolor{blue}{-6.98\%} & \textcolor{red}{-0.31\%} & \textcolor{blue}{-1.3\%} & \textcolor{red}{-0.15\%} & \textcolor{blue}{-4.62\%}
\\
\cmidrule(lr){4-5} \cmidrule(lr){6-7} \cmidrule(lr){8-9} \cmidrule(lr){10-11} \cmidrule(lr){12-13}
& \crossmark &
OD & -10.03&0.11&-9.36&0.09&-10.22&0.07&-9.56&0.08&-10.28&0.11
\\
& \checkmark & 
OD & -10.03&0.11&-9.38&0.08&-10.20&0.07&-9.53&0.10&-10.31&0.10
\\
\cmidrule(lr){4-5} \cmidrule(lr){6-7} \cmidrule(lr){8-9} \cmidrule(lr){10-11} \cmidrule(lr){12-13}
& \multicolumn{2}{c|}{Improv.} &
\textcolor{red}{-0.04\%} & \textcolor{red}{+1.75\%} & \textcolor{red}{-0.2\%} & \textcolor{blue}{-7.0\%} & \textcolor{blue}{+0.13\%} & \textcolor{blue}{-1.62\%} & \textcolor{blue}{+0.31\%} & \textcolor{red}{+21.93\%} & \textcolor{red}{-0.34\%} & \textcolor{blue}{-4.89\%}
\\
\midrule
\rowcolor{lightgray}
\multicolumn{3}{c|}{$\min \Delta$(ID, OD)\%}&
\textcolor{blue}{+0.29\%} & \textcolor{red}{+67.42\%} & \textcolor{blue}{+0.59\%} & \textcolor{blue}{-4.55\%} & \textcolor{blue}{+3.17\%} & \textcolor{red}{+5.02\%} & \textcolor{blue}{+1.77\%} & \textcolor{blue}{-14.37\%} & \textcolor{red}{-0.83\%} & \textcolor{red}{+2.16\%}
\\
\midrule
\multirow{2}{*}{\rotatebox{90}{GLU}} & \crossmark & 
ID & -2.11&0.05&-1.07&0.14&-2.15&0.06&-1.56&0.05&-2.50&0.08
\\
& \crossmark & 
OOD & -1.96&0.11&-1.61&0.10&-1.17&0.12&-1.44&0.06&-2.41&0.09
\\
\midrule
\rowcolor{lightgray}
\multicolumn{3}{c|}{$\min \Delta$(ID, OD)\%}&
\textcolor{blue}{+6.72\%} & \textcolor{red}{+106.76\%} & \textcolor{red}{-50.33\%} & \textcolor{blue}{-29.83\%} & \textcolor{blue}{+45.64\%} & \textcolor{red}{+83.63\%} & \textcolor{blue}{+7.69\%} & \textcolor{red}{+9.24\%} & \textcolor{blue}{+3.33\%} & \textcolor{red}{+6.23\%}
\\
\midrule
\multirow{2}{*}{\rotatebox{90}{LAT}} & \crossmark & 
ID & -25.29&0.36&-10.47&0.25&-52.18&0.42&-20.24&0.30&-26.15&0.33
\\
& \crossmark & 
OOD & -28.75&0.38&-8.80&0.24&-30.19&0.44&-18.19&0.36&-30.08&0.40
\\
\midrule
\rowcolor{lightgray}
\multicolumn{3}{c|}{$\min \Delta$(ID, OD)\%}&
\textcolor{red}{-13.67\%} & \textcolor{red}{+6.5\%} & \textcolor{blue}{+15.89\%} & \textcolor{blue}{-4.01\%} & \textcolor{blue}{+42.14\%} & \textcolor{red}{+4.59\%} & \textcolor{blue}{+10.12\%} & \textcolor{red}{+20.58\%} & \textcolor{red}{-15.03\%} & \textcolor{red}{+20.72\%}
\\
\midrule
\multirow{6}{*}{\rotatebox{90}{NHI}} & \crossmark & 
ID & -10.01&0.12&-9.32&0.11&-10.37&0.10&-9.62&0.09&-10.13&0.11
\\
& \checkmark & 
ID & -10.37&0.07&-9.48&0.21&-10.80&0.08&-9.63&0.07&-10.13&0.11
\\
\cmidrule(lr){4-5} \cmidrule(lr){6-7} \cmidrule(lr){8-9} \cmidrule(lr){10-11} \cmidrule(lr){12-13}
& \multicolumn{2}{c|}{Improv.} &
\textcolor{red}{-3.63\%} & \textcolor{blue}{-37.79\%} & \textcolor{red}{-1.68\%} & \textcolor{red}{+91.12\%} & \textcolor{red}{-4.19\%} & \textcolor{blue}{-21.68\%} & \textcolor{red}{-0.07\%} & \textcolor{blue}{-24.77\%} & \textcolor{red}{-0.01\%} & \textcolor{blue}{-5.46\%}
\\
\cmidrule(lr){4-5} \cmidrule(lr){6-7} \cmidrule(lr){8-9} \cmidrule(lr){10-11} \cmidrule(lr){12-13}
& \crossmark &
OD & -10.08&0.10&-9.26&0.11&-10.18&0.12&-9.49&0.08&-10.20&0.12
\\
& \checkmark & 
OD & -10.21&0.06&-9.36&0.14&-11.10&0.20&-9.58&0.06&-10.19&0.11
\\
\cmidrule(lr){4-5} \cmidrule(lr){6-7} \cmidrule(lr){8-9} \cmidrule(lr){10-11} \cmidrule(lr){12-13}
& \multicolumn{2}{c|}{Improv.} &
\textcolor{red}{-1.3\%} & \textcolor{blue}{-34.64\%} & \textcolor{red}{-1.17\%} & \textcolor{red}{+24.57\%} & \textcolor{red}{-9.0\%} & \textcolor{red}{+66.46\%} & \textcolor{red}{-0.94\%} & \textcolor{blue}{-14.44\%} & \textcolor{blue}{+0.14\%} & \textcolor{blue}{-8.1\%}
\\
\midrule
\rowcolor{lightgray}
\multicolumn{3}{c|}{$\min \Delta$(ID, OD)\%}&
\textcolor{blue}{+1.55\%} & \textcolor{blue}{-16.3\%} & \textcolor{blue}{+1.23\%} & \textcolor{blue}{-34.06\%} & \textcolor{blue}{+1.76\%} & \textcolor{red}{+15.16\%} & \textcolor{blue}{+1.38\%} & \textcolor{blue}{-14.79\%} & \textcolor{red}{-0.55\%} & \textcolor{red}{+4.17\%}
\\
\midrule
\multirow{6}{*}{\rotatebox{90}{TFT}} & \crossmark & 
ID & --&0.16&--&0.07&--&0.23&--&0.07&--&0.07
\\
& \checkmark & 
ID & --&0.30&--&0.16&--&0.25&--&0.08&--&0.06
\\
\cmidrule(lr){4-5} \cmidrule(lr){6-7} \cmidrule(lr){8-9} \cmidrule(lr){10-11} \cmidrule(lr){12-13}
& \multicolumn{2}{c|}{Improv.} &
\textcolor{red}{--\%} & \textcolor{red}{+94.6\%} & \textcolor{red}{--\%} & \textcolor{red}{+114.61\%} & \textcolor{red}{--\%} & \textcolor{red}{+7.57\%} & \textcolor{red}{--\%} & \textcolor{red}{+16.84\%} & \textcolor{red}{--\%} & \textcolor{blue}{-21.55\%}
\\
\cmidrule(lr){4-5} \cmidrule(lr){6-7} \cmidrule(lr){8-9} \cmidrule(lr){10-11} \cmidrule(lr){12-13}
& \crossmark &
OD & --&0.15&--&0.09&--&0.26&--&0.08&--&0.08
\\
& \checkmark & 
OD & --&0.23&--&0.09&--&0.35&--&0.08&--&0.05
\\
\cmidrule(lr){4-5} \cmidrule(lr){6-7} \cmidrule(lr){8-9} \cmidrule(lr){10-11} \cmidrule(lr){12-13}
& \multicolumn{2}{c|}{Improv.} &
\textcolor{red}{--\%} & \textcolor{red}{+57.74\%} & \textcolor{red}{--\%} & \textcolor{red}{+0.35\%} & \textcolor{red}{--\%} & \textcolor{red}{+37.5\%} & \textcolor{red}{--\%} & \textcolor{blue}{-1.64\%} & \textcolor{red}{--\%} & \textcolor{blue}{-35.43\%}
\\
\midrule
\rowcolor{lightgray}
\multicolumn{3}{c|}{$\min \Delta$(ID, OD)\%}&
\textcolor{red}{--\%} & \textcolor{blue}{-22.83\%} & \textcolor{red}{--\%} & \textcolor{blue}{-46.14\%} & \textcolor{red}{--\%} & \textcolor{red}{+10.56\%} & \textcolor{red}{--\%} & \textcolor{red}{+9.88\%} & \textcolor{red}{--\%} & \textcolor{blue}{-9.51\%}
\\
\midrule
\multirow{6}{*}{\rotatebox{90}{TRA}} & \crossmark & 
ID & -9.99&0.23&-9.37&0.21&-10.36&0.12&-9.60&0.13&-10.12&0.11
\\
& \checkmark & 
ID & -10.11&0.21&-9.45&0.31&-10.68&0.18&-9.60&0.10&-10.15&0.11
\\
\cmidrule(lr){4-5} \cmidrule(lr){6-7} \cmidrule(lr){8-9} \cmidrule(lr){10-11} \cmidrule(lr){12-13}
& \multicolumn{2}{c|}{Improv.} &
\textcolor{red}{-1.21\%} & \textcolor{blue}{-6.84\%} & \textcolor{red}{-0.79\%} & \textcolor{red}{+45.69\%} & \textcolor{red}{-3.05\%} & \textcolor{red}{+48.29\%} & \textcolor{blue}{+0.05\%} & \textcolor{blue}{-27.4\%} & \textcolor{red}{-0.34\%} & \textcolor{red}{+0.16\%}
\\
\cmidrule(lr){4-5} \cmidrule(lr){6-7} \cmidrule(lr){8-9} \cmidrule(lr){10-11} \cmidrule(lr){12-13}
& \crossmark &
OD & -9.98&0.19&-9.30&0.22&-10.09&0.14&-9.47&0.15&-10.17&0.12
\\
& \checkmark & 
OD & -10.02&0.11&-9.36&0.22&-10.63&0.25&-9.49&0.08&-10.20&0.12
\\
\cmidrule(lr){4-5} \cmidrule(lr){6-7} \cmidrule(lr){8-9} \cmidrule(lr){10-11} \cmidrule(lr){12-13}
& \multicolumn{2}{c|}{Improv.} &
\textcolor{red}{-0.41\%} & \textcolor{blue}{-43.28\%} & \textcolor{red}{-0.65\%} & \textcolor{red}{+1.0\%} & \textcolor{red}{-5.32\%} & \textcolor{red}{+73.94\%} & \textcolor{red}{-0.16\%} & \textcolor{blue}{-45.49\%} & \textcolor{red}{-0.33\%} & \textcolor{red}{+2.63\%}
\\
\midrule
\rowcolor{lightgray}
\multicolumn{3}{c|}{$\min \Delta$(ID, OD)\%}&
\textcolor{blue}{+0.93\%} & \textcolor{blue}{-47.95\%} & \textcolor{blue}{+0.92\%} & \textcolor{blue}{-28.49\%} & \textcolor{blue}{+2.59\%} & \textcolor{red}{+17.77\%} & \textcolor{blue}{+1.34\%} & \textcolor{blue}{-14.35\%} & \textcolor{red}{-0.53\%} & \textcolor{red}{+8.99\%}
\\

%%%%%%%%%%%%%%%%%%%%%%%%%%%%%%%%%%%%%%%%%%%%%%%%%%%%%%%%%%%%%%%%%%%%%
%%%%%%%%%%%%%%%%%%%%%%%%%%%%%%%%%%%%%%%%%%%%%%%%%%%%%%%%%%%%%%%%%%%%%
%%%%%%%%%%%%%%%%%%%%%%%%%%%%%%%%%%%%%%%%%%%%%%%%%%%%%%%%%%%%%%%%%%%%%
\bottomrule
\end{tabular}}
\end{table}

\subsection{Feature Importance}

{Based on the performance results reported in Tables~\ref{table:results-accuracy} on Task 1 (accuracy) and in Table~\ref{table:results-likelihood}, XGBoost \citep{chen2016xgboost} is the only model that consistently works better with inclusion of extraneous covariates, improves in accuracy on 3 out of 5 datasets and uncertainty quantification on 4 out of 5 datasets. Table\ref{table:feat-importance-xgb} lists the top 10 covariates selected by XGBoost for each dataset. For time-varying features, such as the 36 heart rate observations in Dubosson, the maximum importance across the input length is considered as the feature importance. Below, we provide a discussion on the selected features.

Among features available for all datasets, dynamic time features, such as hour and day of the week, consistently appear in the top 3 important features across all datasets. This could serve as indication that patients tend to adhere to daily routines; therefore, including time features helps the model to predict more accurately. At the same time, patient unique identifier does not appear to be important, only appearing in the top 10 for Broll (Broll only has 7 covariates in total) and Colas. This could be indicative of the fact that differences between patients is explained well by other covariates.

Dynamic physical activity features such as heart rate and blood pressure are only available for Dubosson. Based on the table, we see that medication intake, heart rate and blood pressure metrics, and physical activity measurements are all selected by XGBoost as highly important. 

Demographic and medical record information is not available for Broll and Dubosson. For the rest of the datatsets, we observe medication (e.g. Vitamin D, Lisinopril for Weinstock), disease indicators (e.g. Diabetes T2 for Colas, Osteoporosis for Weinstock), health summary metrics (Body Mass Index for Colas), as well as indices derived from CGM measurements (e.g. J-index \citep{wojcicki1995j}) being selected as highly important.

\begin{table}[h]
{
\caption{Top-10 features with importance weights selected by XGBoost for each dataset.}
% , $p(\cdot|x)$ denotes inclusion of covariates such as demographic information. To compare models with and without covariates, we include the percent improvement on ID and OD test splits. To compare the performance on the ID versus on the OD data, we compute the smallest percent difference between ID and OD for models with and without covariates, denoted as $\min \Delta$(ID, OD)\%
\label{table:feat-importance-xgb}
\centering
\resizebox{\textwidth}{!}{%
\begin{tabular}{cc|cc|cc|cc|cc}
\toprule
\multicolumn{2}{c}{Broll} & \multicolumn{2}{c}{Colas} & \multicolumn{2}{c}{Dubosson} & \multicolumn{2}{c}{Hall} & \multicolumn{2}{c}{Weinstock}  \\ 
\midrule
Covariate &  Importance & Covariate &  Importance &  Covariate &  Importance & Covariate &  Importance & Covariate &  Importance\\
\midrule
Month &    0.001428 & Hour &    0.000634 & Slow Insulin Intake &    0.000625 & Day of week &    0.002322 &  Minute &    0.000444 \\
Day of week &    0.001144 & Day of week &    0.000202 & Hour &    0.000390 & Median CGM &    0.002044 & Day of week &    0.000365 \\
Second &    0.000886 & Glycemia &    0.000133 & heart Rate Variability Index &    0.000359 & J Index of CGM &    0.001808 & Hour &    0.000291 \\
Hour &    0.000768 & Minute &    0.000126 & Body Temperature &    0.000323 & Hour &    0.001786 & Vitamin D &    0.000197 \\
Minute &    0.000410 & Diabetes T2 &    0.000121 & Posture &    0.000296 & Freq. High CGM  &    0.001576 & Year &    0.000194 \\
Patient ID &    0.000072 & Patient ID &    0.000104 & Activity &    0.000292 & Freq. Low CGM &    0.001153 & Erectile dysfunction &    0.000154 \\
Year &    0.000000& \# Follow Up Visits &    0.000093 & Calories &    0.000291 & Minute &    0.001136 & Osteoporosis &    0.000140 \\
& & Body Mass Index (BMI) &    0.000090 & Heart Rate &    0.000197 & \% Pre-Diabetec CGM &    0.001054 & Chronic kidney disease &    0.000140 \\
& & Age &    0.000085 & Blood Pressure &    0.000190 & Coefficient of CGM Variation &    0.000779 & \# of Meter Checks per Day &    0.000137 \\
& & Gender &    0.000070 & Fast Insulin Intake &    0.000140 & Variance of CGM &    0.000706 & Lisinopril &    0.000128 \\
\bottomrule
\end{tabular}}
}
\end{table}

}

\subsection{Stability}

{Reproducible model performance is crucial in the clinical settings. In Table~\ref{table:mae-std}, we report standard deviation of MAE across random data splits. As expected, the smallest datasets (\citet{broll2021interpreting} and \citet{dubosson2018open}) have largest variability. The number of patients in \citet{broll2021interpreting} and \citet{dubosson2018open} is 5 and 9, respectively, thus randomly selecting 1 subject for OD test set has large impact on the model performance as the training set is altered drastically.

Prior works on deep learning has found that initial weights can have large impact on the performance \citep{lecun2002efficient, sutskever2013importance, zhu2022robustness}. Therefore, we re-run each deep learning model 10 times with random initial weights for each data split and report the average. We also report standard deviation of deep learning model results across random model initializations (indicated in parentheses). We find that good initialization indeed matters as we observe that the results differ across re-runs with different starting weights. Such behavior could be undesirable in the clinical settings as the model training cannot be automated. The Transformer is the only robust deep learning model that consistently converges to the same results irrespective of the initial weights, which is reflected in near 0 standard deviation. At the same time, Transformer-based models such as Gluformer and TFT do not exhibit this feature.}

{We include standard errors of each metric: RMSE (Task 1) in Table~\ref{table:rmse-std}, MAE (Task 1) in Table~\ref{table:mae-std}, likelihood (Task 2) in Table~\ref{table:likeli-std}, and calibration in Table~\ref{table:calib-std}. For deep learning models, there are 2 sources of randomness: random data split and model initialization. Therefore, we report two values for standard deviation: one across data splits (averaged over model initializations) and one for model initialization (averaged across data splits). }

\begin{table}[h]
{
\caption{Standard error of MSE across data splits and model random initializations.}
% , $p(\cdot|x)$ denotes inclusion of covariates such as demographic information. To compare models with and without covariates, we include the percent improvement on ID and OD test splits. To compare the performance on the ID versus on the OD data, we compute the smallest percent difference between ID and OD for models with and without covariates, denoted as $\min \Delta$(ID, OD)\%
\label{table:rmse-std}
\centering
\resizebox{\textwidth}{!}{%
\begin{tabular}{ccc|ccccc}
\toprule
& $p(\cdot | x)$ & Data & Broll & Colas & Dubosson & Hall & Weinstock \\
\midrule
\multirow{2}{*}{\rotatebox{90}{ARI}} & \crossmark & 
ID & 110.90 +- 21.95&33.60 +- 0.68&183.11 +- 40.58&74.52 +- 2.25&179.54 +- 3.14
\\
& \crossmark & 
OD & 138.08 +- 52.73&34.97 +- 5.65&351.53 +- 227.93&67.55 +- 25.18&251.97 +- 18.59
\\
\midrule
\multirow{4}{*}{\rotatebox{90}{LIN}} & \crossmark & 
ID & 136.49 +- 13.04&27.70 +- 0.33&145.65 +- 30.12&54.51 +- 1.67&185.04 +- 2.53
\\
& \checkmark & 
ID & 99.04 +- 7.45&30.86 +- 0.71&154.04 +- 32.62&61.45 +- 3.42&179.38 +- 2.56
\\
& \crossmark &
OD & 143.43 +- 50.18&28.41 +- 2.55&246.19 +- 148.20&61.78 +- 17.90&242.59 +- 21.51
\\
& \checkmark & 
OD & 542.88 +- 617.76&30.69 +- 2.55&41255489536.00 +- 82510977335.77&202.15 +- 319.21&245.15 +- 20.05
\\
\midrule
\multirow{4}{*}{\rotatebox{90}{XGB}} & \crossmark & 
ID & 163.83 +- 6.84&41.23 +- 1.08&448.43 +- 28.97&57.45 +- 1.51&185.87 +- 2.80
\\
& \checkmark & 
ID & 192.97 +- 9.24&40.64 +- 3.07&436.29 +- 31.56&64.82 +- 1.99&189.59 +- 2.78
\\
& \crossmark &
OD & 95.32 +- 7.37&38.19 +- 0.79&308.87 +- 21.86&56.15 +- 1.34&236.05 +- 1.27
\\
& \checkmark & 
OD & 93.56 +- 5.14&40.43 +- 3.09&304.16 +- 9.07&67.22 +- 5.03&241.65 +- 3.12
\\
\midrule
\multirow{2}{*}{\rotatebox{90}{GLU}} & \crossmark & 
ID & 201.47 +- 1.24 (20.19)&66.69 +- 1.32 (5.83)&472.51 +- 43.48 (56.73)&59.98 +- 0.47 (1.92)&198.06 +- 4.93 (5.01)
\\
& \crossmark & 
OD & 278.74 +- 6.65 (40.23)&48.10 +- 2.06 (3.13)&551.14 +- 175.34 (56.50)&66.74 +- 6.26 (3.37)&254.05 +- 8.37 (10.83)
\\
\midrule
\multirow{2}{*}{\rotatebox{90}{LAT}} & \crossmark & 
ID & 206.55 +- 18.61 (75.95)&39.38 +- 0.55 (1.36)&405.47 +- 39.58 (68.69)&50.84 +- 0.59 (1.15)&183.46 +- 0.70 (4.84)
\\
& \crossmark & 
OD & 223.91 +- 21.59 (67.04)&31.86 +- 0.70 (1.42)&301.97 +- 104.32 (71.92)&59.38 +- 1.66 (1.57)&226.72 +- 14.59 (7.39)
\\
\midrule
\multirow{4}{*}{\rotatebox{90}{NHI}} & \crossmark & 
ID & 190.20 +- 8.89 (9.07)&35.20 +- 0.52 (0.26)&304.60 +- 63.70 (1.16)&59.02 +- 0.61 (0.27)&176.60 +- 4.41 (1.95)
\\
& \checkmark & 
ID & 262.29 +- 53.94 (26.54)&82.55 +- 21.48 (6.25)&926.26 +- 183.16 (31.93)&66.58 +- 0.73 (1.42)&179.71 +- 2.55 (0.39)
\\
& \crossmark &
OD & 214.26 +- 6.19 (10.83)&32.22 +- 0.06 (0.15)&331.18 +- 128.00 (2.35)&59.97 +- 3.90 (0.21)&210.94 +- 9.26 (2.16)
\\
& \checkmark & 
OD & 245.25 +- 73.67 (37.93)&57.18 +- 13.68 (4.05)&1395.29 +- 755.46 (67.57)&73.77 +- 2.47 (1.95)&207.37 +- 13.01 (0.95)
\\
\midrule
\multirow{4}{*}{\rotatebox{90}{TFT}} & \crossmark & 
ID & 188.64 +- 47.62 (125.97)&31.58 +- 0.79 (1.50)&337.31 +- 6.85 (43.82)&62.66 +- 2.12 (3.63)&205.19 +- 10.63 (13.71)
\\
& \checkmark & 
ID & 215.41 +- 4.50 (32.20)&42.39 +- 0.36 (0.00)&339.65 +- 0.71 (35.06)&70.83 +- 1.80 (3.25)&224.14 +- 6.01 (5.91)
\\
& \crossmark &
OD & 154.46 +- 32.58 (66.26)&30.40 +- 1.41 (1.86)&306.19 +- 58.47 (46.48)&65.95 +- 6.84 (4.04)&232.66 +- 26.29 (14.68)
\\
& \checkmark & 
OD & 175.62 +- 16.08 (20.39)&33.53 +- 1.02 (0.00)&295.58 +- 18.26 (23.85)&79.73 +- 14.18 (5.57)&239.46 +- 17.02 (5.88)
\\
\midrule
\multirow{4}{*}{\rotatebox{90}{TRA}} & \crossmark & 
ID & 228.61 +- 66.99 (0.00)&41.92 +- 3.73 (0.00)&276.33 +- 39.70 (0.00)&62.22 +- 0.99 (0.00)&174.87 +- 13.02 (0.00)
\\
& \checkmark & 
ID & 164.65 +- 12.09 (0.00)&71.18 +- 0.85 (0.00)&752.25 +- 154.38 (0.00)&56.08 +- 1.68 (0.00)&209.02 +- 9.91 (0.00)
\\
& \crossmark &
OD & 197.03 +- 16.72 (0.00)&35.65 +- 1.57 (0.00)&246.66 +- 113.45 (0.00)&66.89 +- 4.68 (0.00)&200.23 +- 24.32 (0.00)
\\
& \checkmark & 
OD & 189.34 +- 0.39 (0.00)&52.66 +- 4.41 (0.00)&1163.43 +- 672.48 (0.00)&54.77 +- 2.65 (0.00)&243.19 +- 1.14 (0.00)
\\
\bottomrule
\end{tabular}}
}
\end{table}

\begin{table}[h]
{
\caption{Standard error of MAE across data splits and model random initializations.}
% , $p(\cdot|x)$ denotes inclusion of covariates such as demographic information. To compare models with and without covariates, we include the percent improvement on ID and OD test splits. To compare the performance on the ID versus on the OD data, we compute the smallest percent difference between ID and OD for models with and without covariates, denoted as $\min \Delta$(ID, OD)\%
\label{table:mae-std}
\centering
\resizebox{\textwidth}{!}{%
\begin{tabular}{ccc|ccccc}
\toprule
& $p(\cdot | x)$ & Data & Broll & Colas & Dubosson & Hall & Weinstock \\
\midrule
\multirow{2}{*}{\rotatebox{90}{ARI}} & \crossmark & 
ID & 8.67 $\pm$ 0.74&4.80 $\pm$ 0.04&11.06 $\pm$ 0.98&7.34 $\pm$ 0.16&11.25 $\pm$ 0.10
\\
& \crossmark & 
OD & 9.71 $\pm$ 1.93&4.87 $\pm$ 0.38&14.58 $\pm$ 4.75&6.97 $\pm$ 1.19&13.34 $\pm$ 0.52
\\
\midrule
\multirow{4}{*}{\rotatebox{90}{LIN}} & \crossmark & 
ID & 9.71 $\pm$ 0.37&4.35 $\pm$ 0.03&9.97 $\pm$ 1.00&6.33 $\pm$ 0.09&11.46 $\pm$ 0.11
\\
& \checkmark & 
ID & 8.41 $\pm$ 0.24&4.60 $\pm$ 0.04&10.03 $\pm$ 1.11&6.66 $\pm$ 0.18&11.34 $\pm$ 0.10
\\
& \crossmark &
OD & 9.83 $\pm$ 1.58&4.41 $\pm$ 0.20&11.90 $\pm$ 3.51&6.62 $\pm$ 0.91&13.09 $\pm$ 0.61
\\
& \checkmark & 
OD & 16.80 $\pm$ 9.45&4.57 $\pm$ 0.19&67548.59 $\pm$ 135072.57&10.02 $\pm$ 6.68&13.16 $\pm$ 0.57
\\
\midrule
\multirow{4}{*}{\rotatebox{90}{XGB}} & \crossmark & 
ID & 11.50 $\pm$ 0.31&5.49 $\pm$ 0.08&19.09 $\pm$ 0.32&6.55 $\pm$ 0.09&11.61 $\pm$ 0.08
\\
& \checkmark & 
ID & 11.87 $\pm$ 0.24&5.46 $\pm$ 0.21&18.55 $\pm$ 0.82&7.02 $\pm$ 0.12&11.77 $\pm$ 0.16
\\
& \crossmark &
OD & 8.72 $\pm$ 0.45&5.32 $\pm$ 0.07&15.42 $\pm$ 0.68&6.52 $\pm$ 0.11&13.04 $\pm$ 0.04
\\
& \checkmark & 
OD & 8.56 $\pm$ 0.30&5.47 $\pm$ 0.18&15.46 $\pm$ 0.35&7.11 $\pm$ 0.26&13.43 $\pm$ 0.13
\\
\midrule
\multirow{2}{*}{\rotatebox{90}{GLU}} & \crossmark & 
ID & 12.55 $\pm$ 0.07 (0.67)&7.12 $\pm$ 0.08 (0.37)&19.40 $\pm$ 1.11 (1.27)&6.69 $\pm$ 0.03 (0.11)&12.09 $\pm$ 0.17 (0.18)
\\
& \crossmark & 
OD & 14.82 $\pm$ 0.26 (1.09)&6.03 $\pm$ 0.13 (0.21)&20.70 $\pm$ 3.58 (1.05)&7.04 $\pm$ 0.33 (0.18)&13.65 $\pm$ 0.22 (0.32)
\\
\midrule
\multirow{2}{*}{\rotatebox{90}{LAT}} & \crossmark & 
ID & 12.32 $\pm$ 0.60 (2.15)&5.37 $\pm$ 0.04 (0.12)&17.88 $\pm$ 0.85 (1.59)&6.11 $\pm$ 0.04 (0.08)&11.45 $\pm$ 0.01 (0.19)
\\
& \crossmark & 
OD & 13.05 $\pm$ 0.59 (1.87)&4.84 $\pm$ 0.05 (0.13)&15.12 $\pm$ 2.76 (1.92)&6.61 $\pm$ 0.12 (0.10)&12.72 $\pm$ 0.40 (0.25)
\\
\midrule
\multirow{4}{*}{\rotatebox{90}{NHI}} & \crossmark & 
ID & 12.07 $\pm$ 0.33 (0.31)&5.04 $\pm$ 0.04 (0.02)&14.79 $\pm$ 1.60 (0.05)&6.57 $\pm$ 0.03 (0.02)&11.21 $\pm$ 0.16 (0.07)
\\
& \checkmark & 
ID & 14.64 $\pm$ 1.57 (0.81)&8.03 $\pm$ 1.40 (0.36)&27.97 $\pm$ 3.06 (0.57)&7.10 $\pm$ 0.03 (0.07)&11.31 $\pm$ 0.09 (0.02)
\\
& \crossmark &
OD & 12.77 $\pm$ 0.18 (0.35)&4.83 $\pm$ 0.01 (0.02)&15.59 $\pm$ 3.33 (0.04)&6.62 $\pm$ 0.19 (0.02)&12.24 $\pm$ 0.25 (0.07)
\\
& \checkmark & 
OD & 14.01 $\pm$ 2.32 (1.03)&6.65 $\pm$ 1.03 (0.27)&33.52 $\pm$ 10.07 (0.83)&7.53 $\pm$ 0.19 (0.11)&12.12 $\pm$ 0.37 (0.03)
\\
\midrule
\multirow{4}{*}{\rotatebox{90}{TFT}} & \crossmark & 
ID & 11.07 $\pm$ 1.17 (2.85)&4.54 $\pm$ 0.05 (0.12)&15.49 $\pm$ 0.05 (1.23)&6.61 $\pm$ 0.12 (0.20)&11.76 $\pm$ 0.35 (0.43)
\\
& \checkmark & 
ID & 12.43 $\pm$ 0.08 (0.89)&5.27 $\pm$ 0.04 (0.00)&15.51 $\pm$ 0.02 (0.83)&7.06 $\pm$ 0.09 (0.19)&12.30 $\pm$ 0.15 (0.17)
\\
& \crossmark &
OD & 10.23 $\pm$ 1.05 (1.91)&4.47 $\pm$ 0.11 (0.15)&14.53 $\pm$ 1.26 (1.11)&6.76 $\pm$ 0.34 (0.21)&12.50 $\pm$ 0.77 (0.45)
\\
& \checkmark & 
OD & 11.17 $\pm$ 0.55 (0.59)&4.68 $\pm$ 0.12 (0.00)&14.43 $\pm$ 0.36 (0.63)&7.44 $\pm$ 0.65 (0.26)&12.67 $\pm$ 0.43 (0.17)
\\
\midrule
\multirow{4}{*}{\rotatebox{90}{TRA}} & \crossmark & 
ID & 13.20 $\pm$ 2.31 (0.00)&5.65 $\pm$ 0.38 (0.00)&14.04 $\pm$ 1.00 (0.00)&6.78 $\pm$ 0.01 (0.00)&11.22 $\pm$ 0.39 (0.00)
\\
& \checkmark & 
ID & 11.27 $\pm$ 0.45 (0.00)&7.77 $\pm$ 0.15 (0.00)&24.40 $\pm$ 2.69 (0.00)&6.42 $\pm$ 0.10 (0.00)&12.61 $\pm$ 0.43 (0.00)
\\
& \crossmark &
OD & 12.28 $\pm$ 0.83 (0.00)&5.24 $\pm$ 0.28 (0.00)&12.98 $\pm$ 2.91 (0.00)&7.07 $\pm$ 0.14 (0.00)&11.91 $\pm$ 0.72 (0.00)
\\
& \checkmark & 
OD & 12.13 $\pm$ 0.03 (0.00)&6.59 $\pm$ 0.36 (0.00)&28.21 $\pm$ 8.72 (0.00)&6.29 $\pm$ 0.12 (0.00)&13.58 $\pm$ 0.06 (0.00)
\\
\bottomrule
\end{tabular}}
}
\end{table}

\begin{table}[h]
{
\caption{Standard error of likelihood across data splits and model random initializations.}
% , $p(\cdot|x)$ denotes inclusion of covariates such as demographic information. To compare models with and without covariates, we include the percent improvement on ID and OD test splits. To compare the performance on the ID versus on the OD data, we compute the smallest percent difference between ID and OD for models with and without covariates, denoted as $\min \Delta$(ID, OD)\%
\label{table:likeli-std}
\centering
\resizebox{\textwidth}{!}{%
\begin{tabular}{ccc|ccccc}
\toprule
& $p(\cdot | x)$ & Data & Broll & Colas & Dubosson & Hall & Weinstock \\
\midrule
\multirow{2}{*}{\rotatebox{90}{ARI}} & \crossmark & 
ID & -9.93 $\pm$ 0.02 & -9.30 $\pm$ 0.05 & -10.47 $\pm$ 0.35 & -9.81 $\pm$ 0.15& -10.21 $\pm$ 0.02
\\
& \crossmark & 
OD & -10.06 $\pm$ 0.05&-9.38 $\pm$ 0.04&-10.44 $\pm$ 0.03&-9.66 $\pm$ 0.56&-10.32 $\pm$ 0.05
\\
\midrule
\multirow{4}{*}{\rotatebox{90}{LIN}} & \crossmark & 
ID & -9.89 $\pm$ 0.01&-9.19 $\pm$ 0.01&-10.10 $\pm$ 0.16&-9.56 $\pm$ 0.03&-10.14 $\pm$ 0.00
\\
& \checkmark & 
ID & -9.87 $\pm$ 0.03&-9.17 $\pm$ 0.01&-10.15 $\pm$ 0.17&-10.30 $\pm$ 1.47&-10.12 $\pm$ 0.00
\\
& \crossmark &
OD & -9.95 $\pm$ 0.14&-9.16 $\pm$ 0.06&-10.11 $\pm$ 0.28&-9.53 $\pm$ 0.17&-10.22 $\pm$ 0.03
\\
& \checkmark & 
OD & -10.24 $\pm$ 0.30&-9.16 $\pm$ 0.06&-12.08 $\pm$ 3.94&-10.42 $\pm$ 1.49&-11.13 $\pm$ 1.83
\\
\midrule
\multirow{4}{*}{\rotatebox{90}{XGB}} & \crossmark & 
ID & -9.94 $\pm$ 0.02&-9.42 $\pm$ 0.01&-10.55 $\pm$ 0.02&-9.68 $\pm$ 0.01&-10.20 $\pm$ 0.00
\\
& \checkmark & 
ID & -10.06 $\pm$ 0.06&-9.40 $\pm$ 0.02&-10.54 $\pm$ 0.01&-9.70 $\pm$ 0.00&-10.21 $\pm$ 0.00
\\
& \crossmark &
OD & -10.03 $\pm$ 0.01&-9.36 $\pm$ 0.01&-10.22 $\pm$ 0.02&-9.56 $\pm$ 0.01&-10.28 $\pm$ 0.00
\\
& \checkmark & 
OD & -10.03 $\pm$ 0.01&-9.38 $\pm$ 0.02&-10.20 $\pm$ 0.01&-9.53 $\pm$ 0.01&-10.31 $\pm$ 0.00
\\
\midrule
\multirow{2}{*}{\rotatebox{90}{GLU}} & \crossmark & 
ID & -2.11 $\pm$ 0.10 (0.24)&-1.07 $\pm$ 0.11 (0.19)&-2.15 $\pm$ 0.01 (0.22)&-1.56 $\pm$ 0.00 (0.10)&-2.50 $\pm$ 0.02 (0.05)
\\
& \crossmark & 
OD & -1.96 $\pm$ 0.08 (0.27)&-1.61 $\pm$ 0.03 (0.12)&-1.17 $\pm$ 1.53 (1.69)&-1.44 $\pm$ 0.12 (0.11)&-2.41 $\pm$ 0.01 (0.05)
\\
\midrule
\multirow{2}{*}{\rotatebox{90}{LAT}} & \crossmark & 
ID & -25.29 $\pm$ 1.96 (5.68)&-10.47 $\pm$ 0.01 (0.16)&-52.18 $\pm$ 2.17 (9.54)&-20.24 $\pm$ 0.39 (0.19)&-26.15 $\pm$ 0.03 (0.07)
\\
& \crossmark & 
OD & -28.75 $\pm$ 2.31 (6.38)&-8.80 $\pm$ 0.25 (0.12)&-30.19 $\pm$ 3.38 (5.20)&-18.19 $\pm$ 0.46 (0.16)&-30.08 $\pm$ 1.49 (0.14)
\\
\midrule
\multirow{4}{*}{\rotatebox{90}{NHI}} & \crossmark & 
ID & -10.01 $\pm$ 0.01 (0.01)&-9.32 $\pm$ 0.01 (0.00)&-10.37 $\pm$ 0.04 (0.00)&-9.62 $\pm$ 0.01 (0.00)&-10.13 $\pm$ 0.00 (0.00)
\\
& \checkmark & 
ID & -10.37 $\pm$ 0.07 (0.05)&-9.48 $\pm$ 0.10 (0.02)&-10.80 $\pm$ 0.01 (0.01)&-9.63 $\pm$ 0.01 (0.01)&-10.13 $\pm$ 0.00 (0.00)
\\
& \crossmark &
OD & -10.08 $\pm$ 0.00 (0.01)&-9.26 $\pm$ 0.01 (0.00)&-10.18 $\pm$ 0.14 (0.00)&-9.49 $\pm$ 0.03 (0.00)&-10.20 $\pm$ 0.03 (0.00)
\\
& \checkmark & 
OD & -10.21 $\pm$ 0.13 (0.04)&-9.36 $\pm$ 0.10 (0.01)&-11.10 $\pm$ 0.51 (0.04)&-9.58 $\pm$ 0.01 (0.01)&-10.19 $\pm$ 0.03 (0.00)
\\
\midrule
\multirow{4}{*}{\rotatebox{90}{TRA}} & \crossmark & 
ID & -9.99 $\pm$ 0.09 (0.00)&-9.37 $\pm$ 0.04 (0.00)&-10.36 $\pm$ 0.04 (0.00)&-9.60 $\pm$ 0.03 (0.00)&-10.12 $\pm$ 0.00 (0.00)
\\
& \checkmark & 
ID & -10.11 $\pm$ 0.11 (0.00)&-9.45 $\pm$ 0.00 (0.00)&-10.68 $\pm$ 0.08 (0.00)&-9.60 $\pm$ 0.00 (0.00)&-10.15 $\pm$ 0.00 (0.00)
\\
& \crossmark &
OD & -9.98 $\pm$ 0.03 (0.00)&-9.30 $\pm$ 0.03 (0.00)&-10.09 $\pm$ 0.06 (0.00)&-9.47 $\pm$ 0.02 (0.00)&-10.17 $\pm$ 0.03 (0.00)
\\
& \checkmark & 
OD & -10.02 $\pm$ 0.01 (0.00)&-9.36 $\pm$ 0.01 (0.00)&-10.63 $\pm$ 0.27 (0.00)&-9.49 $\pm$ 0.02 (0.00)&-10.20 $\pm$ 0.03 (0.00)
\\
\bottomrule
\end{tabular}}
}
\end{table}

\begin{table}[h]
{
\caption{Standard error of calibration error across data splits and model random initializations.}
% , $p(\cdot|x)$ denotes inclusion of covariates such as demographic information. To compare models with and without covariates, we include the percent improvement on ID and OD test splits. To compare the performance on the ID versus on the OD data, we compute the smallest percent difference between ID and OD for models with and without covariates, denoted as $\min \Delta$(ID, OD)\%
\label{table:calib-std}
\centering
\resizebox{\textwidth}{!}{%
\begin{tabular}{ccc|ccccc}
\toprule
& $p(\cdot | x)$ & Data & Broll & Colas & Dubosson & Hall & Weinstock \\
\midrule
\multirow{2}{*}{\rotatebox{90}{ARI}} & \crossmark & 
ID & 0.11 $\pm$ 0.01&0.10 $\pm$ 0.01&0.10 $\pm$ 0.01&0.10 $\pm$ 0.02&0.12 $\pm$ 0.01
\\
& \crossmark & 
OD & 0.07 $\pm$ 0.02&0.08 $\pm$ 0.01&0.08 $\pm$ 0.05&0.07 $\pm$ 0.01&0.12 $\pm$ 0.01
\\
\midrule
\multirow{4}{*}{\rotatebox{90}{LIN}} & \crossmark & 
ID & 0.12 $\pm$ 0.01&0.15 $\pm$ 0.00&0.18 $\pm$ 0.02&0.10 $\pm$ 0.00&0.11 $\pm$ 0.00
\\
& \checkmark & 
ID & 0.13 $\pm$ 0.02&0.19 $\pm$ 0.01&0.21 $\pm$ 0.02&0.19 $\pm$ 0.20&0.11 $\pm$ 0.00
\\
& \crossmark &
OD & 0.15 $\pm$ 0.04&0.15 $\pm$ 0.01&0.17 $\pm$ 0.02&0.10 $\pm$ 0.02&0.11 $\pm$ 0.00
\\
& \checkmark & 
OD & 0.55 $\pm$ 0.39&0.17 $\pm$ 0.02&0.48 $\pm$ 0.58&0.23 $\pm$ 0.18&0.21 $\pm$ 0.20
\\
\midrule
\multirow{4}{*}{\rotatebox{90}{XGB}} & \crossmark & 
ID & 0.07 $\pm$ 0.01&0.10 $\pm$ 0.00&0.07 $\pm$ 0.01&0.09 $\pm$ 0.00&0.11 $\pm$ 0.00
\\
& \checkmark & 
ID & 0.07 $\pm$ 0.01&0.09 $\pm$ 0.01&0.06 $\pm$ 0.01&0.09 $\pm$ 0.00&0.10 $\pm$ 0.00
\\
& \crossmark &
OD & 0.11 $\pm$ 0.01&0.09 $\pm$ 0.00&0.07 $\pm$ 0.01&0.08 $\pm$ 0.01&0.11 $\pm$ 0.00
\\
& \checkmark & 
OD & 0.11 $\pm$ 0.01&0.08 $\pm$ 0.01&0.07 $\pm$ 0.01&0.10 $\pm$ 0.01&0.10 $\pm$ 0.00
\\
\midrule
\multirow{2}{*}{\rotatebox{90}{GLU}} & \crossmark & 
ID & 0.05 $\pm$ 0.01 (0.01)&0.14 $\pm$ 0.01 (0.03)&0.06 $\pm$ 0.00 (0.02)&0.05 $\pm$ 0.00 (0.01)&0.08 $\pm$ 0.00 (0.01)
\\
& \crossmark & 
OD & 0.11 $\pm$ 0.01 (0.03)&0.10 $\pm$ 0.01 (0.02)&0.12 $\pm$ 0.02 (0.05)&0.06 $\pm$ 0.01 (0.01)&0.09 $\pm$ 0.00 (0.01)
\\
\midrule
\multirow{2}{*}{\rotatebox{90}{LAT}} & \crossmark & 
ID & 0.36 $\pm$ 0.03 (0.05)&0.25 $\pm$ 0.01 (0.03)&0.42 $\pm$ 0.02 (0.03)&0.30 $\pm$ 0.01 (0.02)&0.33 $\pm$ 0.01 (0.02)
\\
& \crossmark & 
OD & 0.38 $\pm$ 0.01 (0.05)&0.24 $\pm$ 0.02 (0.03)&0.44 $\pm$ 0.11 (0.08)&0.36 $\pm$ 0.04 (0.03)&0.40 $\pm$ 0.01 (0.03)
\\
\midrule
\multirow{4}{*}{\rotatebox{90}{NHI}} & \crossmark & 
ID & 0.12 $\pm$ 0.02 (0.01)&0.11 $\pm$ 0.00 (0.00)&0.10 $\pm$ 0.00 (0.00)&0.09 $\pm$ 0.01 (0.00)&0.11 $\pm$ 0.00 (0.00)
\\
& \checkmark & 
ID & 0.07 $\pm$ 0.01 (0.01)&0.21 $\pm$ 0.03 (0.04)&0.08 $\pm$ 0.04 (0.02)&0.07 $\pm$ 0.01 (0.00)&0.11 $\pm$ 0.00 (0.00)
\\
& \crossmark &
OD & 0.10 $\pm$ 0.00 (0.01)&0.11 $\pm$ 0.00 (0.00)&0.12 $\pm$ 0.01 (0.00)&0.08 $\pm$ 0.01 (0.00)&0.12 $\pm$ 0.00 (0.00)
\\
& \checkmark & 
OD & 0.06 $\pm$ 0.01 (0.02)&0.14 $\pm$ 0.02 (0.03)&0.20 $\pm$ 0.07 (0.04)&0.06 $\pm$ 0.01 (0.01)&0.11 $\pm$ 0.00 (0.00)
\\
\midrule
\multirow{4}{*}{\rotatebox{90}{TFT}} & \crossmark & 
ID & 0.16 $\pm$ 0.06 (0.08)&0.07 $\pm$ 0.02 (0.03)&0.23 $\pm$ 0.07 (0.10)&0.07 $\pm$ 0.02 (0.02)&0.07 $\pm$ 0.03 (0.03)
\\
& \checkmark & 
ID & 0.30 $\pm$ 0.08 (0.12)&0.16 $\pm$ 0.08 (0.03)&0.25 $\pm$ 0.03 (0.09)&0.08 $\pm$ 0.01 (0.02)&0.06 $\pm$ 0.03 (0.03)
\\
& \crossmark &
OD & 0.15 $\pm$ 0.08 (0.09)&0.09 $\pm$ 0.03 (0.04)&0.26 $\pm$ 0.04 (0.13)&0.08 $\pm$ 0.01 (0.03)&0.08 $\pm$ 0.03 (0.04)
\\
& \checkmark & 
OD & 0.23 $\pm$ 0.05 (0.10)&0.09 $\pm$ 0.07 (0.02)&0.35 $\pm$ 0.04 (0.10)&0.08 $\pm$ 0.01 (0.02)&0.05 $\pm$ 0.02 (0.02)
\\
\midrule
\multirow{4}{*}{\rotatebox{90}{TRA}} & \crossmark & 
ID & 0.23 $\pm$ 0.07 (0.02)&0.21 $\pm$ 0.09 (0.03)&0.12 $\pm$ 0.01 (0.00)&0.13 $\pm$ 0.01 (0.00)&0.11 $\pm$ 0.01 (0.00)
\\
& \checkmark & 
ID & 0.21 $\pm$ 0.05 (0.02)&0.31 $\pm$ 0.07 (0.02)&0.18 $\pm$ 0.04 (0.01)&0.10 $\pm$ 0.00 (0.00)&0.11 $\pm$ 0.00 (0.00)
\\
& \crossmark &
OD & 0.19 $\pm$ 0.03 (0.01)&0.22 $\pm$ 0.11 (0.04)&0.14 $\pm$ 0.03 (0.01)&0.15 $\pm$ 0.01 (0.00)&0.12 $\pm$ 0.00 (0.00)
\\
& \checkmark & 
OD & 0.11 $\pm$ 0.03 (0.01)&0.22 $\pm$ 0.06 (0.02)&0.25 $\pm$ 0.09 (0.03)&0.08 $\pm$ 0.01 (0.00)&0.12 $\pm$ 0.00 (0.00)
\\
\bottomrule
\end{tabular}}
}
\end{table}

\subsection{Daytime versus Nighttime Error Distribution}

\begin{figure}[h]
    \centering
    \includegraphics[width=0.98\textwidth]{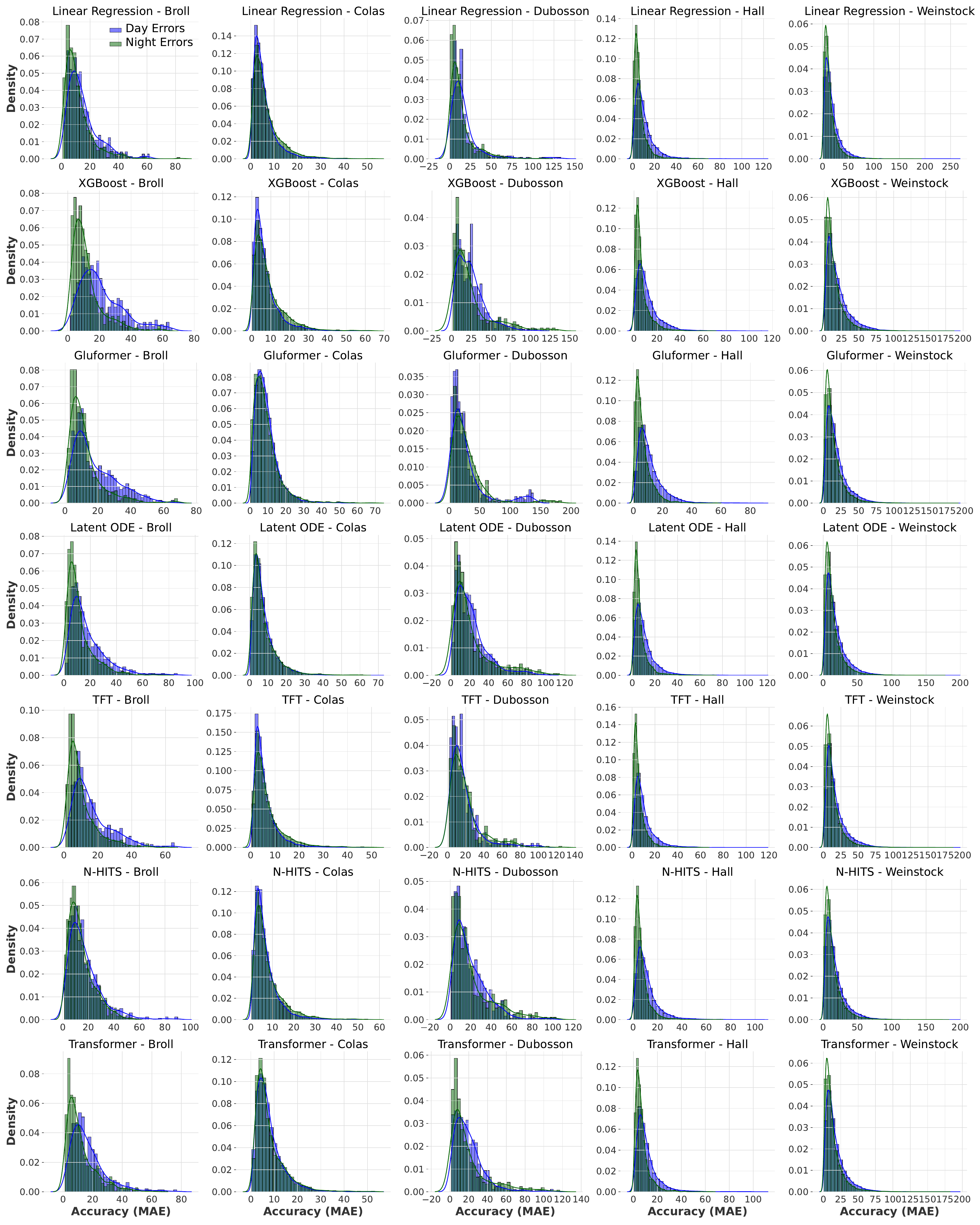}
    \caption{Distribution of daytime (9:00AM to 9:00PM) versus nighttime (9:00PM to 9:00AM) errors (MAE) for models with no covariates on the ID set.}
    \label{fig:nighttime-vs-daytime-all}
\end{figure}

We provide daytime and nighttime error (MAE) distribution for all models and datasets in Figure~\ref{fig:nighttime-vs-daytime-all}. In general, we note that for larger datasets (Colas and Weinstock), the difference in daytime and nighttime error distribution appears smaller. 

\clearpage

\section{Reproducing Results}
\label{appendix:c}

\subsection{Compute Resources}
We conducted all experiments on a single compute node equipped with 4 NVIDIA RTX2080Ti 12GB GPUs. We used Optuna \citep{akiba2019optuna} to tune the hyperparameters of all models except ARIMA and saved the best configurations in the \verb|config/| folder of our repository. 
% (\verb|github.com/IrinaStatsLab/GlucoBench|). 
For ARIMA, we used the native hyperparameter selection algorithm (AutoARIMA) proposed in \citet{hyndmanAutoARIMA2008}. The search grid for each model is available in the \verb|lib/| folder. The training time varied depending on the model and the dataset. We trained all deep learning models using the Adam optimizer for 100 epochs with early stopping that had a patience of 10 epochs. For AutoARIMA, we used the implementation available in \citet{garza2022statsforecast}. For the linear regression, XGBoost \citep{chen2016xgboost}, NHiTS \citep{challuNhits2022n}, TFT \citep{lim2021temporal}, and Transformer \citep{vaswani2017attention}, we used Darts \citep{dartsPackage2022} library. For the Gluformer \citep{sergazinov2022gluformer} and Latent ODE \citep{rubanova2019latent} models, we adapted the original implementation available on GitHub. 

The shallow baselines, such as ARIMA, linear regression, and XGBoost, fit within 10 minutes for all datasets. Among the deep learning models, NHiTS was the fastest to fit, taking less than 2 hours on the largest dataset (Weinstock). Gluformer and Transformer required 6 to 8 hours to fit on Weinstock. Latent ODE and TFT were the slowest to fit, taking 10 to 12 hours on Weinstock on average.

\subsection{{Hyperparameters}}

{In this section, we provide an extensive discussion of hyperparameters, exploring their impact on forecasting models' performance across studied datasets. For each model, we have identified the crucial hyperparameters and their ranges based on the paper where they first appeared. We observe that certain models, such as the Latent ODE and TFT, maintain consistent hyperparameters across datasets. In contrast, models like the Transformer and Gluformer exhibit notable variations. We provide a comprehensive list of the best hyperparameters for each datasets in Table \ref{table:mae-std} and provide intuition below.

\textbf{Linear regression and XGBoost \citep{chen2016xgboost}.} These models are not designed to capture the temporal dependence. Therefore, their hyperparameters change considerably between datasets and do not exhibit any particular patterns. 
For example, the maximum tree depth of XGBoost varies by 67\%, ranging from 6 to 10, while tree regularization remains relatively consistent.

\textbf{Transformer \citep{vaswani2017attention}, TFT \citep{lim2021temporal}, Gluformer \citep{sergazinov2022gluformer}.} Both TFT and Gluformer are based on the Transformer architecture and share most of its hyperparameters. For this set of models, we identify the critical parameters to be the number of attention heads, the dimension of the fully-connected layers (absent for TFT), the dimension of the model (hidden\_size for TFT), and the number of encoder and decoder layers. Intuitively, each attention head captures a salient pattern, while the fully-connected layers and model dimensions control the complexity of the pattern. The encoder and decoder layers allow models to extract more flexible representations. With respect to these parameters, all models exhibit similar behavior. For larger datasets, e.g. Colas, Hall, and Weinstock, we observe the best performance with larger values of the parameters. On the other hand, for smaller datasets, we can achieve best performance with smaller models. 

% {In the Gluformer and TFT models, the dimension of the fully-connected layers and the model dimension respectively show variations of 175\% and 40\%, reflecting their alignment with dataset-specific complexities. Attention heads span from 4 to 12 in the Gluformer and 1 to 4 in both Transformer and TFT.}
 % {The Transformer's encoder and decoder layers range from 1 to 4, reflecting its adaptability in data processing, while the Gluformer maintains a consistent single encoder layer across datasets, focusing on recent data, but varies its decoder layers from 1 to 4, showcasing its forecasting flexibility.}

\textbf{Latent ODE \citep{rubanova2019latent}.} Latent ODE is based on the RNN \citep{sutskever2013importance} architecture. Across all models, Latent ODE is the only one that consistently shows the best performance with the same set of hyperparameter values, which we attribute to its hybrid nature. In Latent ODE, hyperparameters govern the parametric form of the ODE. Therefore, we believe the observed results indicate that Latent ODE is potentially capturing the glucose ODE.
% {This uniformity suggests that its configuration is robust and well-suited to handle diverse data complexities without necessitating excessive hyperparameter fine-tuning.}

\textbf{NHiTS \citep{challuNhits2022n}.} In the case of NHiTS, its authors identify kernel\_sizes as the only critical hyperparameter. This hyperparameter is responsible for the kernel size of the MaxPool operation and essentially controls the sampling rate for the subsequent blocks in the architecture. A larger kernel size leads model to focus more on the low-rate information. Based on our findings, NHiTS selects similar kernel sizes for all datasets, reflecting the fact that all datasets have similar patterns in the frequency domain. 
}

    %{The learning rate, a crucial hyperparameter for most models, demonstrates considerable variation across models and datasets. The classical XGBoost model leans towards higher rates, ranging from 0.5 to 0.7, highlighting its broad data-specific optimization scope. On the other hand, deep models like the TFT and Transformer manifest more conservative rates, specifically 0.0034 and 0.00078 in the Weinstock dataset. While TFT displays consistency with just a 10\% variation, N-HiTS exhibits a dramatic shift, adjusting by up to 200\% between datasets.}

\begin{table}[h]
{
\caption{Best hyperparameters for each model and dataset selected by Optuna \cite{akiba2019optuna}. For models that support covariates, we indicate best hyperparameters with covariates in parantheses.}
% , $p(\cdot|x)$ denotes inclusion of covariates such as demographic information. To compare models with and without covariates, we include the percent improvement on ID and OD test splits. To compare the performance on the ID versus on the OD data, we compute the smallest percent difference between ID and OD for models with and without covariates, denoted as $\min \Delta$(ID, OD)\%
\label{table:mae-std}
\centering
\resizebox{\textwidth}{!}{%
\begin{tabular}{cc|ccccc}
\toprule
 & Hyperparameter & Broll & Colas & Dubosson & Hall & Weinstock \\
\midrule
LIN & 
in\_len & 192.00 (12.00) & 12.00 (12.00) & 12.00 (12.00) & 84.00 (60.00) & 84.00 (84.00) \\
\midrule
\multirow{10}{*}{\rotatebox{90}{XGB}} & 
in\_len & 84.00 (96.00) & 120.00 (144.00) & 168.00 (36.00) & 60.00 (120.00) & 84.00 (96.00) \\
 & lr & 0.51 (0.39) & 0.51 (0.88) & 0.69 (0.65) & 0.52 (0.17) & 0.72 (0.48) \\
 & subsample & 0.90 (0.80) & 0.90 (0.90) & 0.80 (0.80) & 0.90 (0.70) & 0.90 (1.00) \\
 & min\_child\_weight & 2.00 (1.00) & 5.00 (3.00) & 5.00 (2.00) & 3.00 (2.00) & 5.00 (2.00) \\
 & colsample\_bytree & 0.80 (1.00) & 0.90 (0.80) & 0.80 (1.00) & 0.90 (0.90) & 1.00 (0.90) \\
 & max\_depth & 9.00 (8.00) & 7.00 (5.00) & 10.00 (6.00) & 6.00 (6.00) & 10.00 (6.00) \\
 & gamma & 0.50 (1.00) & 0.50 (0.50) & 0.50 (1.50) & 2.00 (1.00) & 0.50 (1.50) \\
 & alpha & 0.12 (0.20) & 0.22 (0.06) & 0.20 (0.15) & 0.10 (0.17) & 0.27 (0.16) \\
 & lambda\_ & 0.09 (0.02) & 0.24 (0.09) & 0.28 (0.09) & 0.13 (0.02) & 0.07 (0.03) \\
 & n\_estimators & 416.00 (288.00) & 352.00 (416.00) & 416.00 (480.00) & 256.00 (320.00) & 416.00 (320.00) \\
\midrule
\multirow{7}{*}{\rotatebox{90}{GLU}} & 
in\_len & 96.00 & 96.00 & 108.00 & 96.00 & 144.00 \\
 & max\_samples\_per\_ts & 100.00 & 150.00 & 100.00 & 200.00 & 100.00 \\
 & d\_model & 512.00 & 384.00 & 384.00 & 384.00 & 512.00 \\
 & n\_heads & 4.00 & 12.00 & 8.00 & 4.00 & 8.00 \\
 & d\_fcn & 512.00 & 512.00 & 1024.00 & 1024.00 & 1408.00 \\
 & num\_enc\_layers & 1.00 & 1.00 & 1.00 & 1.00 & 1.00 \\
 & num\_dec\_layers & 4.00 & 1.00 & 3.00 & 1.00 & 4.00 \\
\midrule
\multirow{8}{*}{\rotatebox{90}{LAT}} & 
in\_len & 48.00 & 48.00 & 48.00 & 48.00 & 48.00 \\
 & max\_samples\_per\_ts & 100.00 & 100.00 & 100.00 & 100.00 & 100.00 \\
 & latents & 20.00 & 20.00 & 20.00 & 20.00 & 20.00 \\
 & rec\_dims & 40.00 & 40.00 & 40.00 & 40.00 & 40.00 \\
 & rec\_layers & 3.00 & 3.00 & 3.00 & 3.00 & 3.00 \\
 & gen\_layers & 3.00 & 3.00 & 3.00 & 3.00 & 3.00 \\
 & units & 100.00 & 100.00 & 100.00 & 100.00 & 100.00 \\
 & gru\_units & 100.00 & 100.00 & 100.00 & 100.00 & 100.00 \\
\midrule
\multirow{7}{*}{\rotatebox{90}{NHI}} & 
in\_len & 96.00 (144.00) & 132.00 (96.00) & 108.00 (120.00) & 144.00 (120.00) & 96.00 (96.00) \\
 & max\_samples\_per\_ts & 50.00 (50.00) & 100.00 (50.00) & 50.00 (50.00) & 100.00 (50.00) & 200.00 (50.00) \\
 & kernel\_sizes & 5.00 (3.00) & 3.00 (3.00) & 3.00 (2.00) & 4.00 (5.00) & 4.00 (3.00) \\
 & dropout & 0.13 (0.09) & 0.18 (0.13) & 0.06 (0.16) & 0.05 (0.19) & 0.13 (0.10) \\
 & lr & 0.00 (0.00) & 0.00 (0.00) & 0.00 (0.00) & 0.00 (0.00) & 0.00 (0.00) \\
 & batch\_size & 64.00 (32.00) & 32.00 (48.00) & 32.00 (48.00) & 48.00 (48.00) & 64.00 (32.00) \\
 & lr\_epochs & 16.00 (10.00) & 2.00 (16.00) & 2.00 (12.00) & 2.00 (4.00) & 16.00 (2.00) \\
\midrule
\multirow{8}{*}{\rotatebox{90}{TFT}} & 
in\_len & 168.00 (96.00) & 132.00 (120.00) & 168.00 (120.00) & 96.00 (132.00) & 132.00 (108.00) \\
 & max\_samples\_per\_ts & 50.00 (50.00) & 200.00 (100.00) & 50.00 (50.00) & 50.00 (50.00) & 200.00 (50.00) \\
 & hidden\_size & 80.00 (80.00) & 256.00 (32.00) & 240.00 (240.00) & 160.00 (64.00) & 96.00 (112.00) \\
 & num\_attention\_heads & 4.00 (3.00) & 3.00 (3.00) & 2.00 (1.00) & 2.00 (3.00) & 3.00 (2.00) \\
 & dropout & 0.13 (0.23) & 0.23 (0.11) & 0.25 (0.24) & 0.13 (0.15) & 0.14 (0.15) \\
 & lr & 0.00 (0.01) & 0.00 (0.00) & 0.00 (0.00) & 0.00 (0.00) & 0.00 (0.00) \\
 & batch\_size & 32.00 (32.00) & 32.00 (32.00) & 64.00 (32.00) & 48.00 (32.00) & 48.00 (48.00) \\
 & max\_grad\_norm & 0.53 (0.03) & 0.98 (0.80) & 1.00 (0.09) & 0.43 (0.66) & 1.00 (0.95) \\
\midrule
\multirow{12}{*}{\rotatebox{90}{TRA}} & 
in\_len & 96.00 (108.00) & 108.00 (120.00) & 108.00 (156.00) & 144.00 (132.00) & 96.00 (96.00) \\
 & max\_samples\_per\_ts & 50.00 (50.00) & 200.00 (200.00) & 50.00 (50.00) & 200.00 (150.00) & 50.00 (50.00) \\
 & d\_model & 96.00 (128.00) & 64.00 (128.00) & 32.00 (64.00) & 64.00 (64.00) & 128.00 (128.00) \\
 & n\_heads & 4.00 (2.00) & 2.00 (4.00) & 2.00 (2.00) & 4.00 (4.00) & 2.00 (4.00) \\
 & num\_encoder\_layers & 4.00 (2.00) & 3.00 (4.00) & 1.00 (2.00) & 1.00 (1.00) & 2.00 (1.00) \\
 & num\_decoder\_layers & 1.00 (2.00) & 3.00 (1.00) & 1.00 (1.00) & 1.00 (3.00) & 4.00 (4.00) \\
 & dim\_feedforward & 448.00 (160.00) & 480.00 (128.00) & 384.00 (384.00) & 96.00 (192.00) & 64.00 (448.00) \\
 & dropout & 0.10 (0.04) & 0.12 (0.20) & 0.04 (0.00) & 0.01 (0.13) & 0.00 (0.19) \\
 & lr & 0.00 (0.00) & 0.00 (0.00) & 0.00 (0.00) & 0.00 (0.00) & 0.00 (0.00) \\
 & batch\_size & 32.00 (32.00) & 32.00 (32.00) & 32.00 (48.00) & 48.00 (48.00) & 32.00 (48.00) \\
 & lr\_epochs & 16.00 (20.00) & 8.00 (18.00) & 6.00 (20.00) & 14.00 (4.00) & 4.00 (4.00) \\
 & max\_grad\_norm & 0.67 (0.89) & 0.83 (0.82) & 0.21 (0.10) & 0.43 (0.23) & 0.42 (0.19) \\
\bottomrule
\end{tabular}}
}
\end{table}

{
\clearpage

\section{Challenges}
\label{appendix:d}

In addressing the challenges associated with the implementation of our predictive models in clinical settings, we recognize three pivotal obstacles. Firstly, the challenge posed by computing power necessitates a strategic refinement of our models to guarantee their effectiveness on devices grappling with resource limitations and potential disruptions in internet connectivity. The delicate balance between the complexity of the model and its real-time relevance emerges as a critical factor, especially within the dynamic contexts of diverse healthcare settings.

Secondly, the challenge of cold starts for new enrolling patients presents a significant hurdle. We acknowledge the importance of devising strategies to initialize and tailor the predictive models for individuals who are newly enrolled in the system. This consideration underscores the need for a dynamic and adaptable framework that ensures the seamless integration of our models into the continuum of patient care.

The third challenge pertains to data privacy and transmission. To address this, our models must either possess on-device training capabilities or facilitate the secure and anonymized transmission of data to external servers. This emphasis on safeguarding patient information aligns with contemporary standards of privacy and ethical considerations, reinforcing the responsible deployment of our models in clinical practice.

}

\bibliography{references}
\bibliographystyle{iclr2024_conference}

\end{document}